\RequirePackage{fix-cm}
\documentclass[11pt]{article}

\usepackage{fullpage}
\usepackage[small,bf]{caption}
\usepackage{amsfonts,dsfont,amssymb,amsbsy,amsmath,paralist,theorem,bm,ifthen,color}
\usepackage[pdfstartview=FitH,bookmarksnumbered,unicode,bookmarksopen=true]{hyperref}
\usepackage{graphicx}
\usepackage{epstopdf}
\usepackage{multirow}
\usepackage{subfigure,relsize}
\usepackage{mathrsfs}
\usepackage{bm}
\usepackage{stfloats}
\usepackage{appendix}
\usepackage{url}
\usepackage[noadjust]{cite}
\usepackage{cases,booktabs}
\usepackage{graphicx,algorithmic,algorithm,relsize}
\usepackage[justification=centering]{caption} % 全局标题居中
\usepackage[table]{xcolor}

\newtheorem{lem}{Lemma}
\newtheorem{prop}{Proposition}

\newtheorem{theo}{Theorem}

\newtheorem{rem}{Remark}
\newtheorem{fact}{Fact}

\newtheorem{assumption}{Assumption}

\newcommand\Jc{\ensuremath{\mathcal{J}}}

\newcommand\yb{\ensuremath{{\bf y}}}

\newcommand\Hb{\ensuremath{{\bf H}}}
\newcommand\hb{\ensuremath{{\bf h}}}
\newcommand\Ab{\ensuremath{{\bf A}}}

\newcommand\Bb{\ensuremath{{\bf B}}}

\newcommand\Db{\ensuremath{{\bf D}}}

\newcommand\Fb{\ensuremath{{\bf F}}}

\newcommand\Gb{\ensuremath{{\bf G}}}

\newcommand\Ib{\ensuremath{{\bf I}}}
\newcommand\Ic{\ensuremath{{\mathcal{I}}}}

\newcommand\Sb{\ensuremath{{\bf S}}}

\newcommand\Xb{\ensuremath{{\bf X}}}
\newcommand\xb{\ensuremath{{\bf x}}}
\newcommand\Yb{\ensuremath{{\bf Y}}}
\newcommand\Ub{\ensuremath{{\bm U}}}
\newcommand\Rb{\ensuremath{{\bf R}}}

\newcommand\Vb{\ensuremath{{\bf V}}}

\newcommand\Wb{\ensuremath{{\bf W}}}

\newcommand\Bbs{\ensuremath{\boldsymbol{\mathcal{B}}}}
\newcommand\Ybs{\ensuremath{\boldsymbol{\mathcal{Y}}}}
\newcommand\Gbs{\ensuremath{\boldsymbol{\mathcal{G}}}}
\newcommand\Hbs{\ensuremath{\boldsymbol{\mathcal{H}}}}
\newcommand\Wbs{\ensuremath{\boldsymbol{\mathcal{W}}}}
\newcommand\Qbs{\ensuremath{\boldsymbol{\mathcal{Q}}}}

\newcommand\Ubs{\ensuremath{\boldsymbol{\mathcal{U}}}}
\newcommand\Sbs{\ensuremath{\boldsymbol{\mathcal{S}}}}
\newcommand\ybs{\ensuremath{\bm{y}}}

\newcommand\hbs{\ensuremath{\bm{h}}}

\newcommand\E{\ensuremath{{\mathbb{E}}}}

\newcommand\diag{\ensuremath{{\rm diag}}}

\newcommand\tr{\ensuremath{{\rm Tr}}}

\newcommand\SNR{\ensuremath{{\rm SNR}}}

\newcommand\Cs{\ensuremath{{\mathbb{C}}}}
\newcommand\Rs{\ensuremath{{\mathbb{R}}}}

\newcommand\Cbb{\ensuremath{{\mathbb{C}}}}

\newcommand\Mset  {\ensuremath{{\mathcal{M}}}}

\newcommand\Hf{\ensuremath{{\mathsf{H}}}}

\graphicspath{{fig/}}

\begin{document}
	%
	% paper title
	% Titles are generally capitalized except for words such as a, an, and, as,
	% at, but, by, for, in, nor, of, on, or, the, to and up, which are usually
	% not capitalized unless they are the first or last word of the title.
	% Linebreaks \\ can be used within to get better formatting as desired.
	% Do not put math or special symbols in the title.
	\title{Low-Complexity Channel Estimation for Massive MIMO Systems with Decentralized Baseband Processing}
	%
	%
	% author names and IEEE memberships
	% note positions of commas and nonbreaking spaces ( ~ ) LaTeX will not break
	% a structure at a ~ so this keeps an author's name from being broken across
	% two lines.
	% use \thanks{} to gain access to the first footnote area
	% a separate \thanks must be used for each paragraph as LaTeX2e's \thanks
	% was not built to handle multiple paragraphs
	%
	
	\author{Yanqing Xu, %\IEEEmembership{Member, IEEE,}
		Bo Wang, 
		Enbin Song, %\IEEEmembership{Senior Member, IEEE,}
		Qingjiang Shi, %\IEEEmembership{Senior Member, IEEE}
		and Tsung-Hui Chang %\IEEEmembership{Senior Member, IEEE,}
		% <-this % stops a space
		\thanks{\smaller[1] The work is supported by Shenzhen Science and Technology Program under Grant No. RCJC20210609104448114, the NSFC, China, under Grant No. 62071409 and by Guangdong Provincial Key Laboratory of Big Data Computing.}
		\thanks{\smaller[1] T.-H. Chang is the corresponding author. }
		\thanks{\smaller[1] Y. Xu and T.-H. Chang are with the School of Science and Engineering, The Chinese University of Hong Kong, Shenzhen, and also with the Shenzhen Research Institute of Big Data, Shenzhen 518172, China (email: xuyanqing@cuhk.edu.cn, tsunghui.chang@ieee.org). }
		\thanks{\smaller[1] Bo Wang is with the Wireless Network RAN Algorithm Department, Xi'an Huawei Technologies Co. Ltd., Xi'an 710000, China (e-mail: wangbo169@huawei.com).}
		\thanks{\smaller[1] E. Song is with the College of Mathematics and School of Aeronautics and Astronautics, Sichuan University, Chengdu, Sichuan 610064, China (email:e.b.song@163.com).}
		\thanks{\smaller[1] Q. Shi is with the School of Software Engineering, Tongji University,
		Shanghai 201804, China, and also with the Shenzhen Research Institute of Big Data, Shenzhen 518172, China (e-mail: shiqj@tongji.edu.cn).}
		\thanks{\smaller[1] This work has been submitted to the IEEE for possible publication. Copyright may be transferred without notice, after which this version may no longer be accessible.} 
	}

	\maketitle

	% As a general rule, do not put math, special symbols or citations
	% in the abstract or keywords.
	\begin{abstract}
		The traditional centralized baseband processing architecture
		is faced with the bottlenecks of high computation
		complexity and excessive fronthaul communication, especially
		when the number of antennas at the base station (BS) is large.
		To cope with these two challenges, the decentralized baseband
		processing (DPB) architecture has been proposed, where the BS
		antennas are partitioned into multiple clusters and each is connected to
		a local baseband unit (BBU). In this paper, we are interested in
		the low-complexity distributed channel estimation (CE) method
		under such DBP architecture, which is rarely studied in the
		literature. The aim is to devise distributed CE algorithms that
		can perform as well as the centralized scheme but with a
		small inter-BBU communication cost. Specifically, based on the
		low-complexity diagonal minimum mean square error channel
		estimator, we propose two distributed CE algorithms, namely
		the aggregate-then-estimate algorithm and the estimate-then-aggregate
		algorithm. In contrast to the existing distributed
		CE algorithm which requires iterative information exchanges
		among the nodes, our algorithms only require one roundtrip
		communication among BBUs. Extensive experiment results
		are presented to demonstrate the advantages of the proposed
		distributed CE algorithms in terms of estimation accuracy,
		inter-BBU communication cost and computation complexity.
	\end{abstract}

		\noindent {\bfseries Keywords} - 
%\begin{IEEEkeywords}
	Massive MIMO, decentralized baseband processing, channel estimation.
%\end{IEEEkeywords}
%\newpage
%	\tableofcontents
%	
%		\newpage
	%	% Note that keywords are not normally used for peerreview papers.
	%	\begin{IEEEkeywords}
	%		IEEE, IEEEtran, journal, \LaTeX, paper, template.
	%	\end{IEEEkeywords}
	%	

	% For peer review papers, you can put extra information on the cover
	% page as needed:
	% \ifCLASSOPTIONpeerreview
	% \begin{center} \bfseries EDICS Category: 3-BBND \end{center}
	% \fi
	%
	% For peerreview papers, this IEEEtran command inserts a page break and
	% creates the second title. It will be ignored for other modes.
%	\IEEEpeerreviewmaketitle

\section{Introduction}
Massive multiple-input multiple-output (MIMO) is an important enabling technique to support the functionalities of 5G and future wireless communication systems \cite{Larsson2014CM}. By deploying a few hundred antennas at the base station (BS), massive MIMO promises to serve multiple mobile users on the same resource block concurrently \cite{Lu2014JSTSP}. 
To fully excavate the potentials of massive MIMO systems, advanced signal processing techniques, such as
precoding and equalization, are developed to exploit the unprecedented spatial degrees of freedom \cite{Gershman2010SPM,Luo2010SPM}.  
However, such algorithm designs rely on the acquisition of accurate channel state information (CSI) of the users, which in practice is estimated via pilot sequence \cite{Li2017TSP,WangTSP2018}. 

Channel estimation (CE) algorithms for massive MIMO systems have been studied extensively, especially the minimum mean square error (MMSE) estimator \cite{Yang2001TCOM,chang-tsp2010,Takano2018TWC,Emil2010TSP}. 
However, despite the high estimation accuracy of the MMSE estimator, it is cursed by both the large storage requirement and the high computation complexity. Specifically, the MMSE estimator relies on the knowledge of the covariance matrix of the channel whose size is quadratic with the number of antennas; also, the MMSE estimator involves large-dimension matrix inversion whose computation complexity is cubic in the number of antennas. 
To reduce both the storage requirement and the computation complexity, \cite{bjornson-jstsp2014} presented a diagonal MMSE (DMMSE) channel estimator which estimates each channel entry individually by assuming a diagonal channel correlation matrix. The DMMSE estimator has a linear complexity with the number of antennas.
Besides, various recent works investigated to exploit the channel sparsity in the angle and delay domains for improving estimation performance while reducing complexity \cite{Gao2020TWC,Fan2018TWC,Kim2019TCOM,Masood2015TSP,gaoTSP2015}.
For example, \cite{Fan2018TWC} proposed an efficient CE algorithm by utilizing the angle-domain sparsity of the millimeter-wave massive MIMO systems, while the sparsity in both the angle and delay domains were exploited in \cite{Kim2019TCOM}.
However, most of the existing CE algorithms are implemented centrally, which requires to pool the signals received by the antennas in a central baseband processor (CBP). 
With the increasing number of antennas, traditional centralized algorithms are faced with the following challenges:  
1) a large fronthaul communication cost between the antennas and baseband processor, and 2) high computation cost due to large-scale signal processing tasks.

Since the above two issues demand a powerful (and expensive) CBP unit, which may no longer be affordable when the antenna size is large, the decentralized baseband processing (DPB) architecture has been proposed recently \cite{Li2017JESTCS}. 
In the DBP architecture, the antennas are divided into several antenna clusters, and each cluster is equipped with an independent (and cheaper) baseband processing unit (BBU) (see Fig. \ref{fig: decentralized network}). 
A naive strategy under the DBP architecture is to let each BBU perform CE based on its locally received signal. 
However, such a fully decentralized scheme would suffer significant performance loss since it neglects the correlation between clusters. Therefore, the idea of the DBP architecture is to leverage advanced  distributed signal processing (SP) techniques to achieve a promising CE performance while having low inter-BBU communication cost and BBU computation complexity.

\subsection{Related Works}
Various distributed/decentralized SP algorithms for the DBP based massive MIMO systems have been investigated in recent years \cite{Li2017JESTCS,Jeon2019TSP,Jeon2017ISIT,Zhang2020TVT,Amiri2022WCL,Rusek-TSP2020,Rusek-TSP2021,Rusek-TSP2022,Zhang2022TVT,Li2019CSSC,Muris-WCL2019,Li2018ACSSC,Croisfelt-ACSSC2021,Kulkarni2021TCS}. 
For instance, the works \cite{Jeon2019TSP,Jeon2017ISIT,Rusek-TSP2021,Rusek-TSP2022,Zhang2022TVT} investigated the uplink equalization algorithms design under the DBP architecture. 
The works \cite{Muris-WCL2019,Li2018ACSSC} studied the decentralized precoding algorithms under the DBP the architecture in the downlink. 
%the interplay between the level of decentralization and the corresponding increase
%%in decentralized processing complexity was investigated in \cite{Rusek-TSP2021}. 
%Recently, considering an extremely large-scale MIMO system, \cite{Croisfelt-ACSSC2021} studied the decentralized equalization problem under the spatially non-stationary channel model.
%While, most of current works considering the star or daisy-chain topology, \cite{Kulkarni2021TCS} investigated the decentralized equalization in a ring network under DBP architecture.
Recently, considering an extremely large-scale MIMO system, \cite{Croisfelt-ACSSC2021} studied the decentralized equalization problem under the spatially non-stationary channel model. While most of the existing works considered the star or daisy-chain topologies, \cite{Kulkarni2021TCS} investigated the decentralized equalization in a ring topology under the DBP architecture.

Surprisingly, there are relatively few works about distributed CE schemes. To the best of our knowledge, paper \cite{AlamTCOM2016} is the only work in the literature that studied the distributed CE problem. 
In particular, \cite{AlamTCOM2016} assumed that the antennas are deployed in a uniform planar array, and each antenna is equipped with a BBU which can only communicate with its neighbors. 
Under this setting, an iterative distributed CE algorithm was proposed. 
However, the method in \cite{AlamTCOM2016} suffers the following drawbacks. 
First, the iterative procedure causes processing delay, and the inter-BBU communication and computation costs increase with the number of iterations. 
Second, in each iteration, the BBUs require to exchange the full-dimension antenna-domain channel estimates, which brings a high inter-BBU communication cost. 
Third, the channel sparsities in the angle and delay domains were not fully exploited.

\subsection{Contributions}
\begin{figure}[t]
	\centering
	\subfigure[star network]{
		\includegraphics[width=0.46\linewidth]{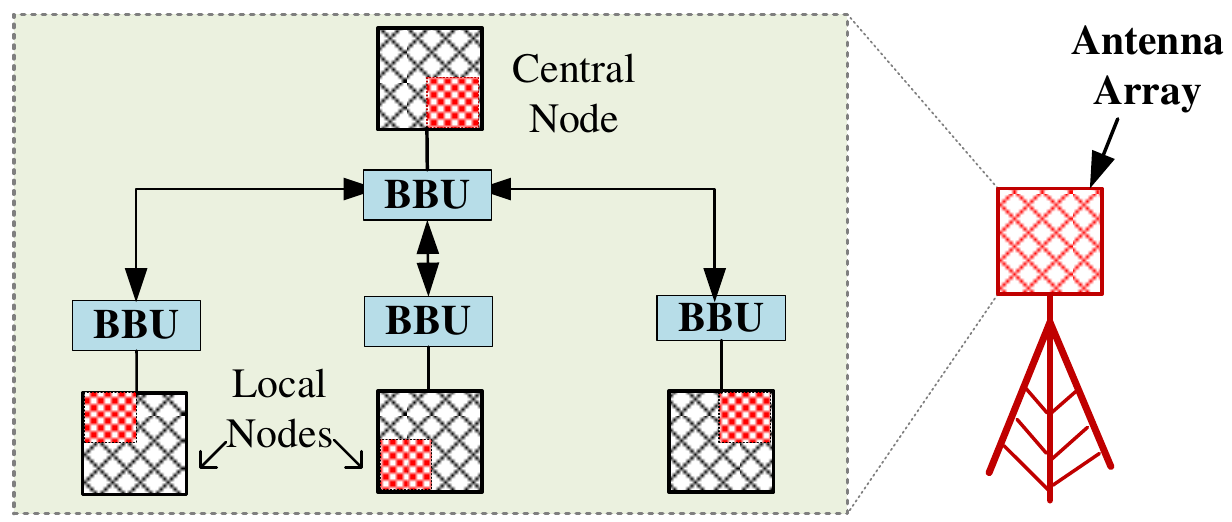}\label{fig: star network}}\\
	\subfigure[daisy-chain network]
	{\includegraphics[width=0.46\linewidth]{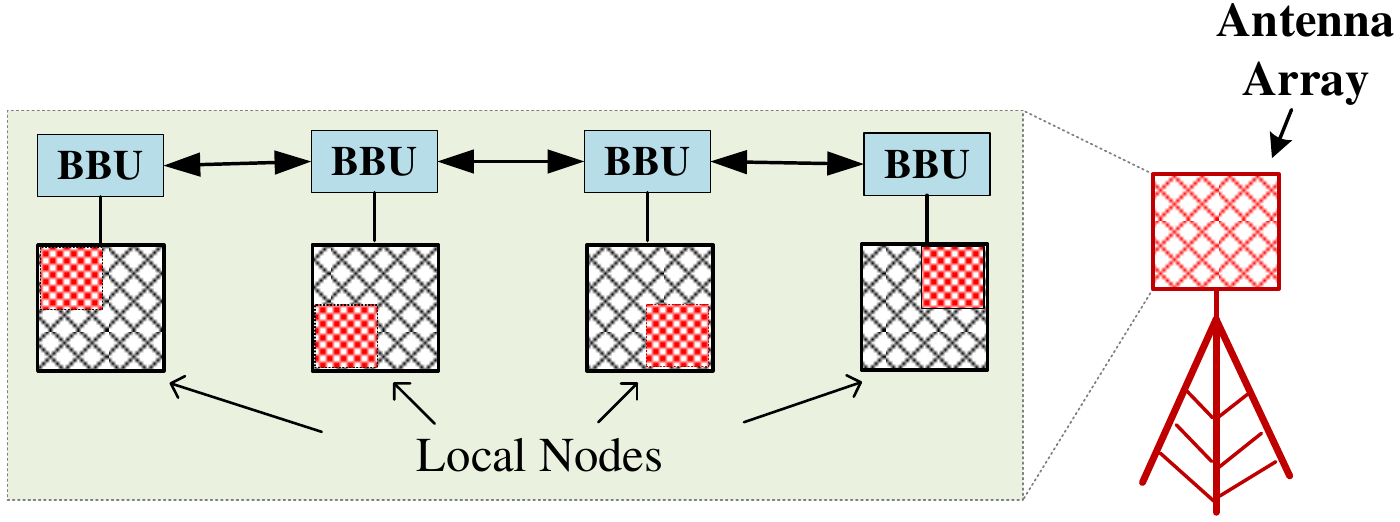}\label{fig: chain network}}
	\caption{Illustrations of the DBP architecture; (a) star network topology and (b) daisy-chain network topology.} 
	\label{fig: decentralized network}
\end{figure}

In this paper, we are interested in the distributed CE algorithm design for the massive MIMO systems under the DBP architecture, as shown in Fig. \ref{fig: decentralized network}. 
Our interest lies in the low-complexity DMMSE estimator and aims to develop distributed CE algorithms that can perform as well as the centralized scheme while maintaining low inter-BBU communication and computation costs. 
Intriguingly, there exist several distinctions between the DMMSE estimator and the conventional MMSE estimator, and they have never been discussed in the literature. 
For example, unlike the MMSE estimator which is equivalent when being applied in the angle-and-delay domain and in the antenna-and-frequency domain, we show that the DMMSE estimator is preferable in the angle-and-delay domain since it can achieve a lower MSE performance. 
Inspired by this insight, we propose two efficient distributed CE algorithms, namely, the aggregate-then-estimate (AGE) and estimate-then-aggregate (EAG) based algorithms, both of which operate in the angle-and-delay domain. 
Compared to \cite{AlamTCOM2016}, the primary merits of the proposed algorithms are that they don't require to iteratively exchange information between BBUs. 
As a result, the computation complexity and inter-BBU communication cost are significantly reduced. 
In addition, the proposed algorithms exploit the channel sparsity in the angle and delay domains and allow flexible control of the tradeoff between inter-BBU communication cost and estimation accuracy. 
The main contributions of this work can be summarized from the following two aspects:
\begin{itemize}
	\item {\bf Novel MSE analyses:} We first prove that, for the DMMSE, estimating the channel coefficients in the angle-and-delay domain achieves a lower MSE than that in the antenna-and-frequency domain. This is particularly true when the channel is sparse in the angle domain, and sparse and dispersive in the delay domain.
	Secondly, we show that  the fully decentralized DMMSE scheme suffers performance loss compared to the centralized one, and the loss increases with the number of clusters. This result suggests the urgent need of developing efficient distributed CE algorithms under the DBP architecture. 
	\item {\bf Efficient distributed CE algorithms:} We propose two novel distributed CE algorithms, namely the AGE-based and EAG-based algorithms. 
	In particular, by exploiting the decomposable structure of the centralized scheme, the AGE-based algorithm only requires one round-trip of information exchange between BBUs. 
	Meanwhile, the channel sparsity in the delay domain is exploited so that only significant coefficients are selected to be exchanged between BBUs. 
	In the EAG-based algorithm, each BBU first estimates its local CSI in the angle-and-delay domain and then sends them for aggregation and refined estimation. 
	Since both the angle- and delay-domain sparsities are exploited, the EAG-based algorithm can achieve a similar performance as the AGE-based algorithm but with an even smaller inter-BBU communication cost. As a tradeoff, the computation complexity of the EAG-based algorithm is slightly higher than the AGE-based algorithm due to the refined estimation at the aggregation node. 
	Both algorithms are shown to perform well in both the star network and the daisy-chain network.
\end{itemize}
Extensive numerical results are presented to validate the efficacy of the proposed algorithms, including MSE, inter-BBU communication cost, and computation complexity. 
Specifically, the overall computation complexities of both algorithms are far smaller than that of the centralized scheme, and the computation loads are evenly distributed among all BBUs. 
Besides, the proposed algorithms can achieve a comparable MSE performance as  the centralized scheme with significantly reduced inter-BBU communication costs in both low and high signal-to-noise ratio (SNR) regimes. 

{\bf Synopsis:} Section \ref{sec: system model and problem formulation} introduces the massive MIMO systems with the DBP architecture and formulates the distributed CE problem. The centralized and fully decentralized baselines are reviewed in Section \ref{sec: baseline review}. Section \ref{sec: proposed algorithms} presents the proposed AGE and EAG-based distributed CE algorithms. The communication costs and computation complexities of the proposed algorithms are summarized in Section \ref{sec: communication and complexity}. Section \ref{sec: numerical result} evaluates the performances of the proposed algorithms by numerical simulations. Finally, the conclusion is drawn in Section \ref{sec: conclusion}.

{\bf Notations:} Column vectors and matrices are denoted by boldfaced lowercase and uppercase letters, e.g., $\xb$ and $\Xb$; $\mathbb{R}^{n \times n}$ and $\mathbb{C}^{n \times n}$ stand for the sets of $n$-dimensional real and complex matrices, respectively. $[\Xb]_j$ and $[\Xb]_{i,j}$ signify the $j$-th column and $(i,j)$-th entry of $\Xb$, respectively.
The superscripts $(\cdot)^\top$, $(\cdot)^\Hf$ and $(\cdot)^*$ describe the transpose, Hermitian, and conjugate operations, respectively. $\tr(\bf X)$ represents the trace of $\Xb$.
 $\diag (\Xb)$ returns a diagonal matrix by setting the off-diagonal entries of $\Xb$ as zero.
$||\xb||_2$ and $||\Xb||_F$ denote the Euclidean norm and Frobenius norm of vector ${\xb}$ and matrix $\Xb$. $\xb = {\rm vec}(\Xb)$ signifies the matrix vectorization.
$\E\{\cdot\}$ represents the statistical expectation operation. $\otimes$ and $\odot$ denote the Kronecker product and Hardamard (element-wise) product, respectively.

\section{System Model and Problem Description} \label{sec: system model and problem formulation}
\subsection{DBP Architecture}
Consider a massive MIMO system where a multi-antenna BS equipped with a uniform linear or planar antenna array communicates with a single-antenna user. The BS possesses a DBP architecture which consists of multiple BBUs. 
The antennas are divided into multiple non-overlapping antenna clusters, and each antenna cluster connects to a dedicated BBU to handle its received signal. 
Each antenna cluster and its BBU are viewed as a local node. Based on the way of information exchange, we consider two different distributed architectures, i.e., the star network and daisy-chain network, as shown in Fig. \ref{fig: decentralized network}.
In the star network, there exists a central node, and the other nodes can only communicate with the central/aggregation node. 
While, in the daisy-chain network, each local node can only communicate with its neighbors. 

\subsection{Signal Model and CE Problem}
In this work, we focus on the uplink CE, where the user transmits predefined pilot signals over $N_C$ subcarriers to the BS equipped with $N_R$ antennas. Without loss of generality, we assume that each antenna cluster consists of $N_r$ antennas and $N_r = \frac{N_R}{M}$, where $M$ is the number of clusters. 
For simplicity, we assume that the pilot matrix is identity.
Then, the received antenna-and-frequency-domain signal at the $m$-th node is given by
\begin{align}
\Yb_m &=\Hb_m +\Wb_m, m\in \mathcal{M},\label{eqn: received signal local node}
\end{align}
where $\mathcal{M}\triangleq\{1,...,M\}$, $\Hb_m \in \Cs^{N_r \times N_C}$ and $\Wb_m$ are the antenna-and-frequency-domain channel and noise at node $m$, respectively. We assume that each column of $\Wb_m$ follows $\mathcal{CN}(0,\sigma_w^2\Ib_{N_r})$ and is independent with other columns. 

The idea of CE is to suppress or remove $\Wb_m$ from the received signal $\Yb_m$ to recover $\Hb_m$. 
Under the DBP architecture, the CE can be implemented in a fully decentralized way where each antenna cluster estimates its local channel solely with its received signal. 
Also, the CE can be carried out in a centralized way by pooling all the information of antenna clusters and estimating the entire channel in the central node. Then the corresponding channel is feedback to the local nodes.
%The local nodes can use their local channels for further signal processing algorithms, such as, equalization and precoding designs.
In the next section, we first review the centralized and fully decentralized CE schemes as the two benchmark methods. Then, we propose two novel distributed CE algorithms in Section \ref{sec: proposed algorithms}.

\section{Centralized and Fully Decentralized Schemes} \label{sec: baseline review}

\subsection{Centralized CE Algorithms}
For centralized CE, the central node collects the antenna-and-frequency-domain signals $\Yb_m,m\in \mathcal{M},$ from all local nodes.
By stacking $\Yb_m$s' together, the received signal at the central node can be written as
\begin{align}\label{eqn: received signal central node antenna frequency domain channel}
	\Yb =\Hb +\Wb \in \Cs^{N_R \times N_C}, 
\end{align}
where $\Yb = [\Yb_1^{\Hf},\Yb_2^{\Hf},\ldots,\Yb_M^{\Hf}]^{\Hf}$, $\Hb = [\Hb_1^\Hf,\Hb_2^\Hf,\ldots,\Hb_M^\Hf]^\Hf$, and $\Wb = [\Wb_1^\Hf,\Wb_2^\Hf,\ldots,\Wb_M^\Hf]^\Hf$. Note that the channel matrix can also be represented by
\begin{align}\label{eqn: received signal central node angle delay domain channel}
	\Hb =\Fb_{N_R} \Hbs \Fb_{N_C}, 
\end{align}
where $\Hbs$ is the angle-and-delay-domain channel matrix, $\Fb_{N_R}$ and $\Fb_{N_C}$ are the discrete Fourier transformation (DFT) matrices of dimension $N_R \times N_R$ and $N_C \times N_C$, respectively.

In what follows, we review two MMSE based centralized CE algorithms, i.e., the full MMSE algorithm, and the diagonal MMSE (DMMSE) algorithm.

{\bf 1) Full MMSE algorithm \cite{Yang2001TCOM}:}
By the MMSE criterion,
the \textit{antenna-and-frequency-domain} channel is estimated by solving 
%the following optimization problem 
\begin{align} \label{eq:mmse antenna frequency domain channel estimation}
	\min_{\Ub} ~&\E \left\{||\Ub {\yb} - {\hb}||_2^2\right\},
\end{align}
where $\yb = {\rm vec}(\Yb)$, ${\hb} = {\rm vec}({\Hb})$, and $\Ub \in \Cbb^{N_RN_C \times N_RN_C}$ is the MMSE estimator.
The optimal $\Ub$ is given by
\begin{align} \label{eqn: antenna frequency domain soft win central}
	\Ub = \Rb_{\hb} \left(\Rb_{\hb} + \sigma_w^2 \Ib_{N_RN_C}\right)^{-1},
\end{align}
where $\Rb_{\hb} = \mathbb E \{\hb \hb^\Hf\}$ is the antenna-and-frequency-domain channel covariance matrix. Then, the antenna-and-frequency-domain channel estimate is given by
\begin{align}\label{eqn: mmse antenna frequency domain channel estimate}
	\widehat{\hb} = \Rb_{\hb} \left(\Rb_{\hb} + \sigma_w^2 \Ib_{N_RN_C}\right)^{-1} \yb.
\end{align}

The channel can also be estimated from the \textit{angle-and-delay-domain}. Define $\ybs$ and $\hbs$ as
\begin{subequations}
	\begin{align}\label{eq:yh_tilde}
		\ybs &= {\rm vec}(\Ybs) = \left(\Fb_{N_C}^* \otimes \Fb_{N_R}^\Hf \right){\yb},\\
		\hbs &= {\rm vec}(\Hbs) = \left(\Fb_{N_C}^* \otimes \Fb_{N_R}^\Hf \right){\hb},
	\end{align}
\end{subequations}
where $\Ybs = \Fb_{N_R}^\Hf \Yb \Fb_{N_C}^\Hf$ is the received signal in the angle-and-delay domain.
The corresponding channel estimate can be obtained by solving
\begin{align} \label{eq:mmse angle delay domain channel estimation}
	\min_{\Ubs} ~&\E \left\{||\Ubs {\ybs} - {\hbs}||_2^2\right\}.
\end{align}
Analogous to \eqref{eqn: antenna frequency domain soft win central}, the optimal $\Ubs$ can be written as
\begin{align} \label{eqn: angle delay domain soft win central}
 \Ubs = \Rb_{\hbs} \left(\Rb_{\hbs} + \sigma_w^2 \Ib_{N_RN_C}\right)^{-1},
\end{align}
where $\Rb_{\hbs} = \mathbb E \{\hbs \hbs^\Hf\}$ is the angle-and-delay domain channel covariance matrix. Then, the corresponding channel estimate reads 
\begin{align}\label{eqn: mmse angle delay domain channel estimate}
\widehat{\hbs} = \Rb_{\hbs} \left(\Rb_{\hbs} + \sigma_w^2 \Ib_{N_RN_C}\right)^{-1} \ybs.
\end{align}
%where $\Rb_{\hbs} = \mathbb E \{\hbs \hbs^\Hf\}$ is the angle-and-delay domain channel covariance matrix.
For the estimation of $\hb$ and $\hbs$, it is easy to verify the following fact is true. 
\begin{fact}
	By the full MMSE criterion, estimating $\hb$ is equivalent to estimating $\hbs$. In particular, $\widehat \hbs = \left(\Fb_{N_C}^* \otimes \Fb_{N_R}^\Hf \right) \widehat\hb$. Thus, the MSEs for estimating $\hb$ and $\hbs$ are the same.
\end{fact}

One can see from \eqref{eqn: mmse antenna frequency domain channel estimate} and \eqref{eqn: mmse angle delay domain channel estimate} that the full MMSE algorithm requires to calculate the inverse of a matrix with dimension of $N_RN_C \times N_RN_C$, which has a computation complexity order of $\mathcal{O}(N_R^3N_C^3)$. Besides, the required storage for $\Ub$ and $\Ubs$ increases  quadratically with the number of antennas. 

{\bf 2) DMMSE algorithm \cite{bjornson-jstsp2014}:} 
To reduce the computation complexity, one can ignore the inter-antenna and inter-frequency correlations and estimate each channel coefficient individually. 
The associated problem can be formulated as 
\begin{align}
	\min_{\substack{\Vb ~\!\rm is\!~diagonal} } ~\E\left\{ \|\Vb \yb - \hb\|_2^2\right\}\label{eqn: diagonal mmse central}
\end{align}
where 
$\Vb \in \Cs^{N_RN_C \times N_RN_C}$ is the associated (diagonal) estimator to be optimized.
It is easy to check that the optimal $\Vb$ is given by
\begin{align}
	\Vb = \diag(\Rb_{\hb})\diag^{-1}(\Rb_{\hb}+ \sigma_w^2\Ib_{N_RN_C}).
\end{align}
Then the antenna-and-frequency-domain channel estimate can be written as 
\begin{align} \label{eqn: antenna-and-frequency-domain channel estimate in each path}
	\widehat{\hb} = \diag(\Rb_{\hb})\diag^{-1}(\Rb_{\hb}+ \sigma_w^2\Ib_{N_RN_C}) \ybs.
\end{align}
By rearranging the entries on the diagonal of $\Vb$ as a matrix $\Sb \in \Cs^{N_R \times N_C}$, the DMMSE based CE can be viewed as a soft windowing operation on the received signal, and the channel estimate can be written as
\begin{align} \label{eqn: antenna frequency channel estimate of central diagonal mmse} 
	\widehat{\Hb} = \Sb \odot \Yb,
\end{align}
where the $(i,j)$-th entry of $\Sb$ is given by
\begin{align}\label{eqn: antenna frequency channel soft windowing matrix}
	[\Sb]_{i,j} = \frac{[\Rb_{\Hb}]_{i,j}}{[\Rb_{\Hb}]_{i,j} \!+\! \sigma_w^2},i\in \mathcal{N_R},j\in \mathcal{N_C},
\end{align} 
where $\Rb_{\Hb} \triangleq \E\{\Hb \odot \Hb^*\} \in \Rs^{N_R \times N_C}$ is the power profile of $\Hb$, $\mathcal{N_R} \triangleq \{1,2,...,N_R\}$ and $\mathcal{N_C} \triangleq \{1,2,...,N_C\}$.
The associated MSE is given by
\begin{align} \label{eqn: mse antenna frequency domain channel central}
	{\rm MSE}^{\rm af} = \sum_{i=1}^{N_R}\sum_{j=1}^{N_C} \frac{[\Rb_{\Hb}]_{i,j}\sigma_w^2}{[\Rb_{\Hb}]_{i,j} + \sigma_w^2},
\end{align}
Similarly, the angle-and-delay domain channel is given by 
\begin{align} \label{eqn: angle delay channel estimate of central diagonal mmse}
	\widehat{\Hbs} = \Sbs \odot \Ybs,
\end{align}
where $\Sbs \in \Cs^{N_R\times N_C}$, and the $(i,j)$-th entry is given by
\begin{align} \label{eqn: angle delay channel soft windowing matrix}
	[\Sbs]_{i,j} = \frac{[\Rb_{\Hbs}]_{i,j}}{[\Rb_{\Hbs}]_{i,j} \!+\! \sigma_w^2}, i\in \mathcal{N_R},j\in \mathcal{N_C},
\end{align} 
where $\Rb_{\Hbs} \triangleq \E\{\Hbs \odot \Hbs^*\} \in \Rs^{N_R \times N_C}$ is the power profile of $\Hbs$.
Then, the associated MSE is given by
\begin{align} \label{eqn: mse angle delay domain channel central}
	{\rm MSE}^{\rm c} = \sum_{i=1}^{N_R}\sum_{j=1}^{N_C} \frac{[\Rb_{\Hbs}]_{i,j}\sigma_w^2}{[\Rb_{\Hbs}]_{i,j} + \sigma_w^2}.
\end{align}

\begin{figure}[t]
	\centering
	\includegraphics[width=0.6\linewidth]{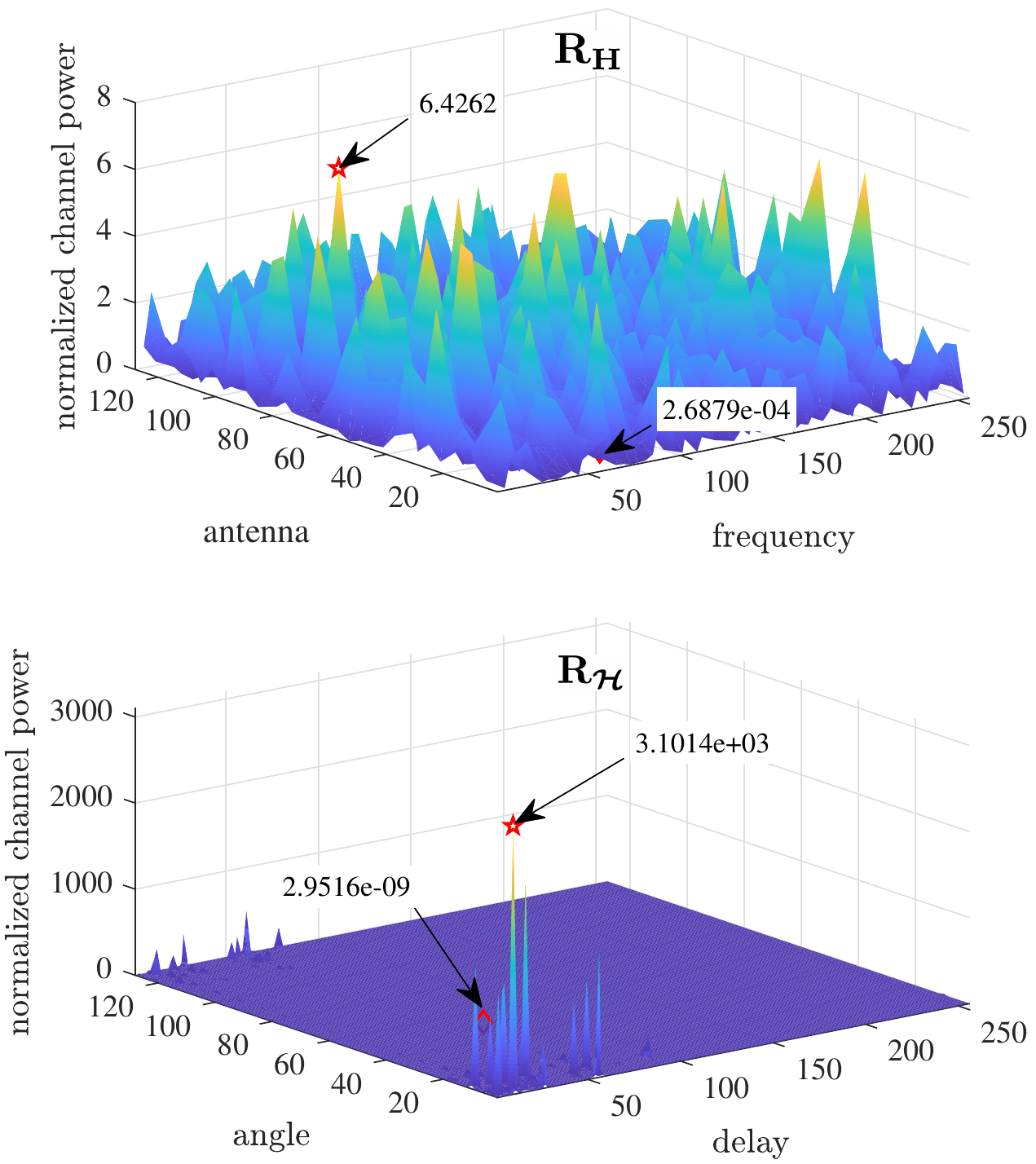}\\
	\caption{Illustration of the power profiles of the MIMO channel in the antenna-and-frequency-domain (upper) and in the angle-and-delay-domain (lower), where the respective maximum and minimum values are marked.}
	\label{fig: channel sparsity} 
\end{figure}

It is intriguing to note that, different from the full MMSE algorithm, estimating $\Hbs$ is not equivalent to estimating $\Hb$ for the DMMSE algorithm. 
The reason is two-fold. First, \eqref{eqn: antenna frequency channel estimate of central diagonal mmse} and \eqref{eqn: angle delay channel estimate of central diagonal mmse} are not linearly related. Second, the power profiles of the angle-and-delay-domain channel and the antenna-and-frequency-domain channel, i.e., $\Rb_{\Hbs}$ and $\Rb_{\Hb}$, are dramatically distinct. 
In Fig. \ref{fig: channel sparsity}, we compare the normalized $\Rb_{\Hbs}$ and $\Rb_{\Hb}$, where the channel was generated by the ``3GPP-38.901-UMa-NLOS'' model in {\tt QuaDRiGa} with $N_R = 128$, $N_C = 256$ and bandwidth equals to $15$MHz \cite{quadriga}. 
One can see that most of the channel power concentrates in a limited number of significant angles and paths only, because the channel is sparse in the angle domain and sparse
%\footnote{It should be pointed out that the angle-and-delay-domain channel is not strictly sparse due to limited samplings by the IDFT transformations.} 
and dispersive in the delay domain due to a limited number of scatters in practical wireless environments.
Based on the above observations, we make the following assumption:
\begin{assumption} \label{assump: comparison of the ad and af channel values}
	The power profiles of $\Hbs$ and $\Hb$, i.e., $\Rb_{\Hbs}$ and $\Rb_{\Hb}$, satisfy that
	\begin{subequations}
		\begin{align}
			\max_{i\in \mathcal{N_R},j\in \mathcal{N_C}} \left[\Rb_{\Hbs}\right]_{i,j} &\geq \max_{i\in \mathcal{N_R},j\in \mathcal{N_C}} \left[\Rb_{\Hb}\right]_{i,j}, \label{eqn: maximum value assumption}\\
			\min_{i\in \mathcal{N_R},j\in \mathcal{N_C}} \left[\Rb_{\Hbs}\right]_{i,j} &\leq \min_{i\in \mathcal{N_R},j\in \mathcal{N_C}} \left[\Rb_{\Hb}\right]_{i,j}. \label{eqn: minimum value assumption}
		\end{align}
	\end{subequations}
\end{assumption}
Then, we have the following theorem.
\begin{theo} \label{theo: mses of diagonal mmse algorithm}
	Suppose Assumption \ref{assump: comparison of the ad and af channel values} hold. For the DMMSE algorithm, we have
\begin{align}
	{\rm MSE}^{\rm c} \leq {\rm MSE}^{\rm af}.
\end{align}
\end{theo}

%\begin{IEEEproof}
	{\emph Proof:} The proof is relegated to Appendix \ref{appd : proof of theorem I}. \hfill $\blacksquare$
%	\end{IEEEproof}

From the proof of Theorem \ref{theo: mses of diagonal mmse algorithm}, one can see that ${\rm MSE}^{\rm c}$ can be strictly smaller than ${\rm MSE}^{\rm af}$ if at least one of \eqref{eqn: maximum value assumption} and \eqref{eqn: minimum value assumption} holds with strict inequality.
Actually, our numerical results show that \eqref{eqn: maximum value assumption} and \eqref{eqn: minimum value assumption} always hold with strict inequality and thus ${\rm MSE}^{\rm c}$ is strictly smaller than ${\rm MSE}^{\rm af}$.

The computation complexity order of the DMMSE is $\mathcal{O}(N_RN_C)$ and also the required storage is only linear with the number of antennas. Considering these advantages of the DMMSE algorithms, we take it as the baseline, and study decentralized/distributed CE algorithms based on it. In the next subsection, let us first consider the naive fully decentralized scheme.

\subsection{Fully Decentralized CE Algorithm}
In the fully decentralized (FD) scheme, the $m$-th node estimates its own channel by solely using the locally received signal $\Yb_m \in \Cbb^{N_r\times N_C},$ without exchanging any information with other nodes.
Following the same idea as that in the centralized DMMSE algorithm, the channel estimate of each antenna cluster by the FD DMMSE algorithm is given by 
\begin{align} \label{eqn: channel estimate local}
	\widehat{\Hbs}_m = \Sbs_m \odot \Ybs_m, \in \Cs^{N_r\times N_C},
\end{align}
where $\Ybs_m = \Fb_{N_r}^\Hf \Yb_m \Fb_{N_C}^\Hf$ is the received angle-and-delay-domain signal of cluster $m$, $\Sbs_m \in \Cs^{N_r \times N_C}$ is given by
\begin{align}
	[\Sbs_m]_{i,j} =\frac{[\Rb_{\Hbs_m}]_{i,j}}{[\Rb_{\Hbs_m}]_{i,j} + \sigma_w^2},m\in \mathcal{M},i\in\mathcal{N}_r,j\in\mathcal{N_C}.
\end{align}
Here, $\Hbs_m = \Fb_{N_r}^\Hf \Hb_m \Fb_{N_C}^\Hf$ is the local angle-and-delay-domain channel of cluster $m$, $\Rb_{\Hbs_m} \triangleq \E\{\Hbs_m \odot \Hbs_m^*\}$ signifies the power profile of
$\Hbs_m$, and  $\mathcal{N}_r \triangleq \{1,...,N_r\}$.
Then, the aggregated MSE of all $M$ clusters is given by
	\begin{align} \label{eqn: mse angle delay domain channel local}
		{\rm MSE}^{\rm FD} & = \sum_{m=1}^M \sum_{i=1}^{N_r} \sum_{j=1}^{N_C} \frac{[\Rb_{\Hbs_m}]_{i,j} \sigma_w^2 }{[\Rb_{\Hbs_m}]_{i,j} + \sigma_w^2}.
	\end{align}

Since the channel of each node is estimated locally, there is no inter-BBU communication cost. It is easy to understand that the FD scheme would suffer from performance loss compared to the centralized scheme. However, analytically proving this fact is by no means trivial since \eqref{eqn: mse angle delay domain channel local} is not a simple degeneration of \eqref{eqn: mse angle delay domain channel central}.  We overcome this based on an argument similar to that for proving Theorem \ref{theo: mses of diagonal mmse algorithm}.

\begin{figure}[t]
	\centering
	\includegraphics[width=0.52\linewidth]{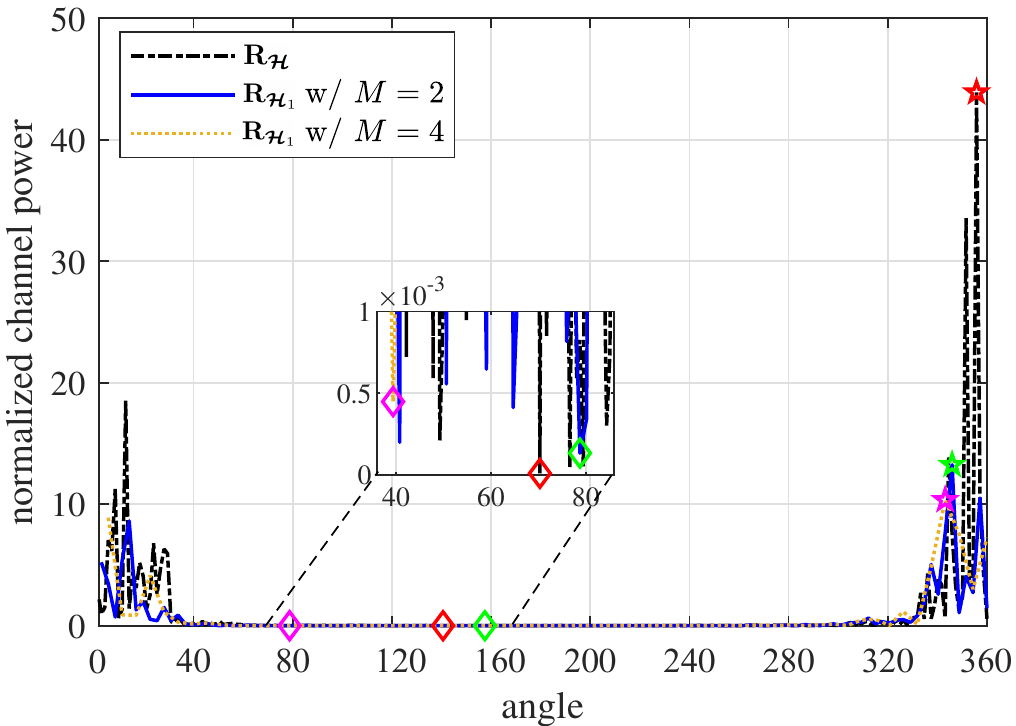}\\
	\caption{Comparison of the angle-domain channel power profiles of the entire channel $\Hbs$ and local channel $\Hbs_1$ of the first antenna cluster, where the ``red'', ``green'' and ``purple'' diamonds signify the minimum values of $\Rb_{\Hbs}$, $\Rb_{\Hbs_1}$ with $M=2$ and $M=4$, respectively.} 
	\label{fig: sparsity comparison of the centralized and FD scheme}
\end{figure}

Let us compare the power profiles of $\Hbs$ and $\Hbs_m$, i.e., $\Rb_{\Hbs}$ and $\Rb_{\Hbs_m}$. 
We assume $N_R=256$, $N_C=1$ and generated the channel by the ``3GPP-38.901-UMa-NLOS'' model in {\tt QuaDRiGa}. In Fig. \ref{fig: sparsity comparison of the centralized and FD scheme}, we plot the entry values of $\Rb_{\Hbs}$ and $\Rb_{\Hbs_1}$. As one can see from this figure, with the increase of $M$, the maximum (minimal) value of $\Rb_{\Hbs_1}$ decreases (increases).
%The reason is that the spatial resolution of $\hbs = \Fb_{N_R} \hb$ is larger than that of $\hbs_m = \Fb_{N_r}\hb_m$. 
The reason is that, with larger $M$, each cluster has fewer antennas which not only makes $\Rb_{\Hbs_m}$ have less energy but also lower resolution in the angle domain. 
Based on this observations, we make the following assumption.\vspace{-0mm}
\begin{assumption} \label{assump: comparison of the central and fd channel values}
	The power profiles of $\Hbs$ and $\Hbs_m$, i.e., $\Rb_{\Hbs}$ and $\Rb_{\Hbs_m} $, satisfy that
	\begin{subequations}
		\begin{align}
			\max_{i\in \mathcal{N_R},j\in \mathcal{N_C}} \left[\Rb_{\Hbs}\right]_{i,j} &\geq \max_{i\in \mathcal{N}_r,j\in \mathcal{N_C}} \left[\Rb_{\Hbs_m}\right]_{i,j}, \forall m \in \mathcal{M}, \label{eqn: maximum value assumption 2}\\
			\min_{i\in \mathcal{N_R},j\in \mathcal{N_C}} \left[\Rb_{\Hbs}\right]_{i,j} &\leq \min_{i\in \mathcal{N}_r,j\in \mathcal{N_C}} \left[\Rb_{\Hbs_m}\right]_{i,j}, \forall m \in \mathcal{M}. \label{eqn: minimum value assumption 2}
		\end{align}
	\end{subequations}
\end{assumption}

Then, we have the following corollary.
\begin{theo} \label{theo : mse central and fd}
	Suppose Assumption \ref{assump: comparison of the central and fd channel values} hold.
	For any $2 \leq M \leq N_R$, it is always true that 
	\begin{align}
		{\rm MSE}^{\rm c} \leq {\rm MSE}^{\rm FD}.
	\end{align}
\end{theo}
%\begin{IEEEproof}
	{\emph Proof:} The proof of  Theorem \ref{theo : mse central and fd} follows the same idea as that in the proof of Theorem \ref{theo: mses of diagonal mmse algorithm} by using Assumption \ref{assump: comparison of the central and fd channel values}, Lemma \ref{lem : concavity}, and the fact that $\sum_{i=1}^{N_R}\sum_{j=1}^{N_C} [\Rb_{\Hbs}]_{i,j} = \sum_{m=1}^M \sum_{i=1}^{N_r} \sum_{j=1}^{N_C}[\Rb_{\Hbs_{m}}]_{i,j}$. Due to limited space, the details are omitted here. \hfill $\blacksquare$
%\end{IEEEproof}

\begin{figure}[t]
	\centering
	\includegraphics[width=0.52\linewidth]{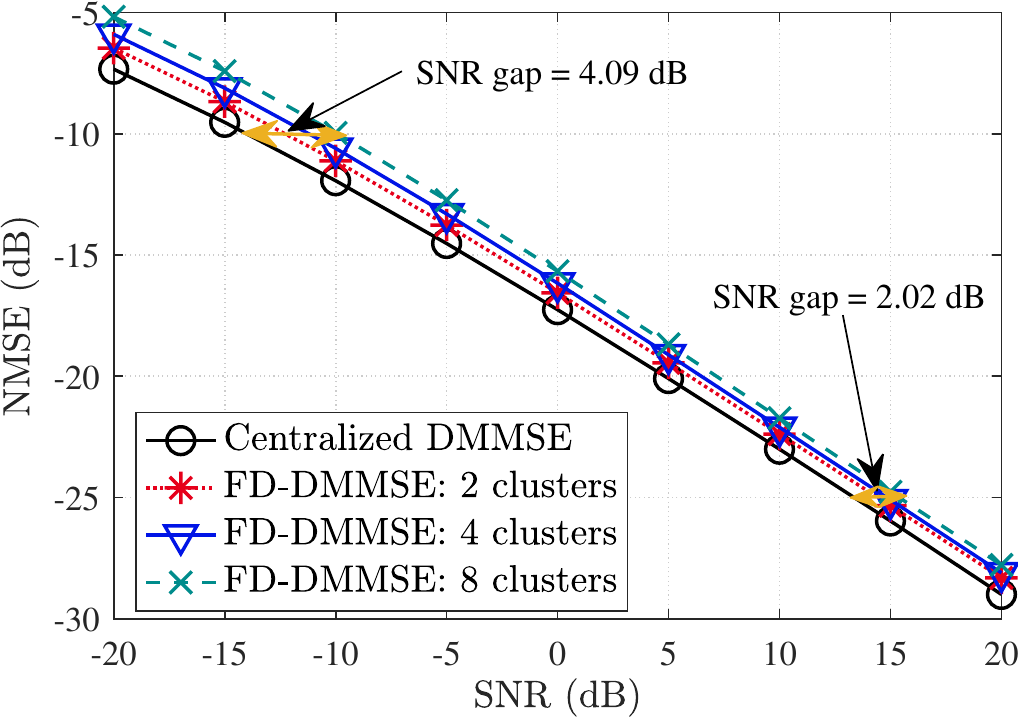}\\
	\caption{Normalized mean square error (NMSE) comparisons of the centralized DMMSE scheme and the FD-DMMSE scheme with different number of clusters, where the NMSE is defined by ${\rm NMSE} = \|\Hb - \widehat \Hb\|_F^2/\|\Hb\|_F^2$.}
	\label{fig: nmse baseline}
\end{figure}

A numerical example to validate Theorem \ref{theo : mse central and fd} is given in Fig. \ref{fig: nmse baseline}. One can see that the centralized algorithm can strictly outperform the FD algorithm. Meanwhile, the NMSE gap between the centralized DMMSE and FD-DMMSE algorithms increases with the number of antenna clusters. 
This also corroborates with Theorem \ref{theo : mse central and fd} since, by observing Fig. \ref{fig: sparsity comparison of the centralized and FD scheme}, the right hand side (RHS) values in \eqref{eqn: maximum value assumption 2} and \eqref{eqn: minimum value assumption 2} decrease and increase respectively with increasing $M$.
%The reason is that with the increase of $M$, the maximum (minimum) value of $\Rb_{\Hbs_m}, \forall m\in \mathcal{M}$, is decreased (increased) due to low spatial resolution as shown in Fig. \ref{fig: sparsity comparison of the centralized and FD scheme}.

\section{Proposed Distributed CE Algorithms} \label{sec: proposed algorithms}
Considering the degraded estimation accuracy of the FD scheme, we are interested in designing distributed CE algorithms which can achieve comparable estimation accuracy to the centralized scheme while having a low inter-BBU communication cost. 
%Remind that Fig. \ref{fig: channel sparsity} shows that the wireless channel in practical environment is sparse in both angle and delay domains. 
%This indicates that most of the channel power is decentralized in limited number of significant angles and paths. 
%Motivated by this, one can design distribute algorithms where the nodes only exchange several significant channel entries to approach the centralized solution while keeping manageable inter-BBU communication cost.
%
In this section, two distributed CE algorithms, namely, the AGE based and the EAG based algorithms, are presented which have the desired merits.

%Specifically, by benefiting the delay-domain sparsity and the decomposable structure of the centralized algorithm, the AGE-based algorithm can approach to the centralized algorithm with quite cheap inter-BBU communication and computation complexity.
%While by exploring both the angle and delay domain sparsity, the EAG-based algorithm achieves better performance compared to the AGE-based algorithm but with a slightly increased computation complexity.
%The details of the proposed algorithms will be described in the following of this section. Let's first introduce the AGE-based decentralized CE algorithm.

\subsection{AGE-Based Distributed Algorithm}
Recall the centralized algorithm in \eqref{eqn: angle delay channel estimate of central diagonal mmse} where the angle-and-delay-domain channel estimate is given by
\begin{align}
	\widehat\Hbs
	= \Sbs \odot\left(\Fb_{N_R}^\Hf \Yb \Fb_{N_C}^\Hf\right).  \label{eqn: central soft windowing agg}
\end{align}
By partitioning $\Fb_{N_R}$ horizontally into $M$ submatrices, i.e., $\Fb_{N_R}=[\Fb_1^\Hf, \Fb_2^\Hf, \ldots,\Fb_M^\Hf]^\Hf$ with $\Fb_m \in \Cs^{N_r \times N_R},m\in \mathcal{M}$, \eqref{eqn: central soft windowing agg} can be written as
\begin{align}
	\widehat\Hbs
	&= \Sbs \odot \left(\sum\nolimits_{m=1}^M\Fb_m^\Hf \Yb_m \Fb_{N_C}^\Hf\right). \label{eqn: central soft windowing agg2}
\end{align}
Given the decomposable structure in \eqref{eqn: central soft windowing agg2} and the delay domain sparsity of wireless channel, we attempt to design the (sparse) aggregation based distributed CE algorithm. In particular, in the AGE-based scheme, the local antenna-delay-domain received signal $\Yb_m \Fb_{N_C}^{\Hf}$ is divided into two parts by hard windowing. The part with larger power values is to be sent to the central node and aggregated with the information from the other nodes for centralized CE, and the remaining part with smaller power values is used for local CE. By this, one can realize flexible tradeoff between estimation accuracy, computation complexity and inter-BBU communication cost.

{\bf \indent 1) Processing at local nodes: }
Let us illustrate the scheme by considering the star network. 
To exploit the delay-domain sparsity, we apply a local windowing matrix $\Db_m \in \mathbb{R}^{N_r\times N_C}$ to $\Yb_m \Fb_{N_C}^{\Hf}$,  that is, 
%Denote $\Db_m \in \mathbb{R}^{N_r\times N_C}$ as a local hard windowing matrix of cluster $m$. By exploiting the delay-domain sparsity, each cluster $m$ obtains a hard windowed signal
\begin{align}
	\bar \Yb_m &= \Db_m \odot \Yb_m \Fb_{N_C}^\Hf  \in \mathbb{C}^{N_r \times N_C}, m\in \mathcal{M},
	\label{eqn: aggregation local windowing}
\end{align}
where $[\Db_m]_{i,j} \in \{0,1\},i\in \mathcal{N_R},j\in \mathcal{N_C}$ are used to select prominent elements in $\Yb_m \Fb_{N_C}^\Hf$. In particular, the $j$-th column of $\Db_m$, i.e., $[\Db_m]_j$, is determined by
\begin{align}\label{eq:column sparsity agg dl}
	[\Db_m]_{i,j} = 
	\begin{cases}
		1, &{\rm if}~ \|[\Yb_m \Fb_{N_C}^\Hf]_j\|^2 \geq \eta N_r \sigma_w^2,\\
		0, &{\rm otherwise},
	\end{cases}
\end{align}
where $\eta$ is a threshold parameter to control the sparsity of $\Db_m$. In particular, a larger $\eta$ results in a sparser $\Db_m$. 
%Notice that, due to the randomness of the noise, the delay-domain sparsity patterns of $\bar \Yb_m$'s maynot be the same.

\begin{rem}
	Comparing to the element-by-element windowing, the advantages of the column-wise windowing are two-fold. 
	First, in the massive MIMO system, the channel is spatially stationary and the delay-domain properties of the channel for all the antennas in a cluster are approximately the same and thus the antenna-delay-domain channel matrix is column-wise sparse \cite{gaoTSP2015,Gong2017TVT}. Second, for the column-wise sparse signal, local nodes only need to upload the values and corresponding indices of the non-zero columns, and therefore the inter-BBU communication costs can be reduced.
\end{rem}

By hard windowing, the antenna-delay domain signal, $\Yb_m \Fb_{N_C}^{\Hf}$, at node $m$ is divided into two parts. One is given by \eqref{eqn: aggregation local windowing}, and the other part, $\bar \Yb_m^{(l)}$, is given by
\begin{align}
	\bar \Yb_m^{(l)} = \Yb_m \Fb_{N_C}^\Hf - {\bar \Yb_m} \in \mathbb{C}^{N_r \times N_C}, m \in \Mset. \label{eqn: aggregation local windowing 2}
\end{align}
%which is used to estimate the corresponding channel locally. 
The local node $m$ can use $\bar \Yb_m^{(l)}$ to obtain a local DMMSE estimate in the antenna-delay domain, which we denote as $\widehat \Gb_m^{(l)} \in \Cs^{N_r\times N_C}$. We also denote $\mathcal{I}_m^{(l)}$ as the index set of the non-zero columns of $\widehat \Gb_m^{(l)}$.

{\bf \indent 2) Processing at central node: }
%After hard windowing, $\bar \Yb_m$'s are to the central node. 
%Then, the central node first transform $\bar\Yb_m$'s to their angle-and-delay domain by oversampling and then aggregates the angle-delay signal to obtain
The windowed signal $\bar \Yb_m$s' are sent to the central node, and the aggregated angle-delay domain signal is given by
\begin{subequations}
\begin{align}
	\bar \Ybs & = \sum\nolimits_{m=1}^M \Fb_m^\Hf {\bar \Yb}_m 	\label{eqn: aggregation central windowing}\\
	& = \Fb_{N_R}^\Hf \bar \Yb \\
	& = \Fb_{N_R}^\Hf ({\bar \Hb} + {\bar \Wb}) \\
%	& = \sum\nolimits_{m=1}^M \Fb_m^\Hf {\bar \Hb}_m  + \sum\nolimits_{m=1}^M \Fb_m^\Hf {\bar \Wb}_m \\
%	& = \Fb_{N_R}^\Hf \bar\Hb + \Fb_{N_R}^\Hf \bar\Wb\\
	& \triangleq \bar \Hbs + \bar \Wbs	
	\in \mathbb{C}^{N_R \times N_C},	
\end{align}
\end{subequations}
where $\bar\Yb = [\bar\Yb_1^\Hf,...,\bar\Yb_M^\Hf]^\Hf$, $\bar\Hb = [\bar\Hb_1^\Hf,...,\bar\Hb_M^\Hf]^\Hf$, $\bar\Wb = [\bar\Wb_1^\Hf,...,\bar\Wb_M^\Hf]^\Hf$, ${\bar \Hb}_m = \Db_m \odot \Hb_m$, and ${\bar \Wb}_m = \Db_m \odot \Wb_m, \forall m \in \Mset$. Each column of $\bar \Wbs$ is independent with other columns, and its $j$-th column
follows $\mathcal{CN}(0,\lambda_j\Ib_{N_r})$ where $\lambda_j = \frac{\bar M}{M} \sigma_w^2$. Here, $\bar M$ denotes the number of clusters that upload nonzero $[\bar \Yb_m]_j, \forall m \in \Mset$.

Then, by following \eqref{eqn: central soft windowing agg2}, the central node estimates the overall angle-and-delay domain channel based on $\bar\Ybs$.
Specifically, denote $\Ic$ as the index set of the non-zero columns of $\bar \Ybs$. The $j$-th column of the centralized channel estimate is given by
\begin{align}\label{eq: dl central estimation}
	[\widehat\Gbs]_j = [\bar\Sbs]_j \odot [\bar\Ybs]_j, \forall j \in \Ic,
\end{align}
where $[\bar\Sbs]_{j} = \frac{[\Rb_{\bar\Hbs}]_{i,j}}{[\Rb_{\bar\Hbs}]_{i,j} + \lambda_j}$ and $\Rb_{\bar\Hbs} = \E\{\bar\Hbs \odot \bar\Hbs^*\}$. While, for $j \notin \Ic$, $[\widehat\Gbs]_{i,j} = 0$. The corresponding MSE of estimating $[\widehat\Gbs]_j$ is given by
\begin{align}\label{eqn: mse central age algorithm}
	{\rm MSE}_j^{\rm c,AGE} = \sum_{i=1}^{N_R}\frac{[\Rb_{\bar\Hbs}]_{i,j}\lambda_j}{[\Rb_{\bar\Hbs}]_{i,j} + \lambda_j}, \forall j \in \Ic.
\end{align}
Due to the oversampling on ${\bar \Yb}_m$ by $\Fb_m^\Hf, \forall m\!\in\! \Mset,$ in \eqref{eqn: aggregation central windowing}, one can conclude that ${\rm MSE}_j^{\rm c,AGE}$ is no larger than that incurred by the FD scheme as per Theorem \ref{theo : mse central and fd}.

The angle-and-delay-domain channel $\widehat\Gbs$ should be sent back to the local nodes. To reduce the inter-BBU communication cost, we transform $\widehat\Gbs$  to its antenna-and-delay domain by
\begin{align} \label{eq: dl central angle to antenna}
	\widehat{\Gb} &= \Fb_{N_R}\widehat\Gbs 
	= [\widehat{\Gb}_1^\Hf, \widehat{\Gb}_2^\Hf, \ldots, \widehat{\Gb}_M^\Hf]^\Hf \in \mathbb{C}^{N_R \times N_C}.
\end{align}
Then, the antenna-and-delay-domain channel estimate {$\widehat{\Gb}_m \in \Cs^{N_r \times N_C}$}, $m\in \mathcal{M}$, are sent to the corresponding local nodes. 
Note that, by \eqref{eq: dl central angle to antenna}, the delay-domain sparsity patterns of {$\widehat{\Gb}_m$}s' are the same as $\widehat{\Gb}$.

{\bf \indent 3) Post-processing at local nodes: }
Note that different local nodes may upload different columns (paths)  of $\Yb_m \Fb_{N_C}^\Hf$ to the central node. Consequently, the non-zero columns of $\widehat{\Gb}_{m}$ and $\widehat{\Gb}^{(l)}_{m}$ may be partially overlapped, which indicates that $\Ic \cap \Ic_m^{(l)} \neq \emptyset$.
To fully utilize all the information, node $m$ can determine the $j$-th column of the antenna-and-delay-domain channel estimate by
	\begin{align}\label{eqn: anntenn delay domain channel of AGE algorithm}
	[{\widetilde \Ab}_m]_j\!=\! 
	\begin{cases}
		[\widehat{\Gb}_m]_j, & \forall j \in \bar\Ic_m ,\\
		[\widehat{\Gb}^{(l)}_m]_j, & \forall j \in \bar\Ic_m^{(l)},\\
		\alpha [\widehat{\Gb}_m]_j \!+\! (1\!-\!\alpha) [\widehat{\Gb}^{(l)}_m]_j,\!\!\!&
		{\rm otherwise},
	\end{cases}
\end{align}
where 
%\begin{align*}
{\small $\bar \Ic_m = \left\{j|j \in \Ic\setminus\Ic_m^{(l)}\right\}$, $\bar \Ic_m^{(l)} = \left\{j|j \in \Ic_m^{(l)}\setminus\Ic\right\},\forall m \in \Mset,$}
%\end{align*}
and $0\leq \alpha \leq 1$ is the convex combination coefficient which can be chosen empirically.

Finally, the antenna-and-frequency-domain channel estimate of each local node is given by
\begin{align}\label{eq: dl agg complete channel estiamte}
	\widehat \Hb_m = {\widetilde \Ab}_{m} \Fb_{N_C} ,m\in \mathcal{M}.
\end{align}
The complete antenna-and-frequency-domain channel estimate is $\widehat\Hb = [\widehat\Hb_1^\Hf,\widehat\Hb_2^\Hf,...,\widehat\Hb_M^\Hf]^\Hf$  and its corresponding MSE is given by ${\rm MSE^{AGE} = \E \{\|\widehat\Hb - \Hb\|^2\}}$. The details of the AGE-based algorithm are summarized in Algorithm \ref{alg:agg based CE}. 

\begin{algorithm}[!tb] \small
	\caption{AGE-based algorithm for distributed CE}
	\begin{algorithmic}[1]
		\REQUIRE Received antenna-and-frequency-domain signal $\Yb$, noise power $\sigma_w^2$, number of clusters $M$, and threshold parameter $\eta$. \\%\vspace{1mm}
		{\bf 1) Processing at local nodes:}
		\STATE {Each node computes the hard-windowed signal by \eqref{eqn: aggregation local windowing}.}
		\STATE {Node $m,\forall m \in \Mset$, uses ${\bar \Yb}^{(l)}_m$ to estimate  ${\widehat{  \Gb}}^{(l)}_m$.}\\%\vspace{1mm}
		{\bf 2) Processing at central node:}
		\STATE {The central node first transforms the received antenna-delay signal to its angle-and-delay domain by oversampling and then aggregates them by \eqref{eqn: aggregation central windowing}.}
		\STATE {The central node performs the centralized estimation by \eqref{eq: dl central estimation}.}
		\STATE {The central node transforms the estimated channel to its antenna-delay domain by \eqref{eq: dl central angle to antenna} and then sends it to the local nodes.}\\%\vspace{1mm}
		{\bf 3) Post-processing at local nodes:}%\vspace{0mm}
		\STATE {Each node determines its antenna-delay-domain channel by \eqref{eqn: anntenn delay domain channel of AGE algorithm}, and obtains its antenna-and-frequency-domain channel by \eqref{eq: dl agg complete channel estiamte}.}
		\ENSURE {The estimated local channels: {\smaller$\widehat \Hb_m$}$,m\in \mathcal{M}.$}
	\end{algorithmic}\label{alg:agg based CE} 
\end{algorithm}
%
%It should be emphasized that although the removed columns by hard windowing in \eqref{eqn: aggregation local windowing 2} are with small values, it is still important to exploit them to produce a high-accuracy channel estimate. Simulation results to validate this is given in Fig. \ref{fig: mse with or without local information} in Sec. \ref{sec: numerical result}.
%
%\begin{figure*}[t]
%	\centering
%	\includegraphics[width=0.70\linewidth]{information_exchange_stationary_dl_agg}\\
%	\caption{ Schematic illustrations of the information exchange processes of the AGE algorithm where ``CN'' denotes the central node and ``LN'' represents a local node.}
%	\label{fig:information_exchange_stationary_dl_agg}
%\end{figure*}
%
From \eqref{eqn: anntenn delay domain channel of AGE algorithm}, one can see that the MSE of the AGE algorithm comes from three aspects: the first is from the central estimate $[\widehat{\Gb}_m]_j, j \in \bar\Ic_m$; the second is for locally estimating  $[\widehat{\Gb}^{(l)}_m]_j,  j \in \bar\Ic_m^{(l)}$, and the last is their combination  $\alpha [\widehat{\Gb}_m]_j + (1-\alpha) [\widehat{\Gb}^{(l)}_m]_j, j \in  \Ic\cap\Ic_m^{(l)} $. 
The MSE of AGE-based algorithm is given in the following proposition. 
\begin{prop} \label{prop : nmse aggregation}
	The MSE of the AGE-based distributed CE algorithm satisfies
	\begin{align}
		{\rm MSE}^{\rm AGE} \leq& \sum_{j \in \bar\Ic_m}{\rm MSE}_j^{\rm c,AGE} + \sum_{j \in \bar\Ic_m^{(l)}}{\rm MSE}_j^{\rm FD} + \sum_{j \in \Ic\cap\Ic_m^{(l)}}\left(\alpha {\rm MSE}_j^{\rm c,AGE} + (1-\alpha){\rm MSE}_j^{\rm FD}\right),
	\end{align}
where
${\rm MSE}_j^{\rm FD} = \sum_{m=1}^M \sum_{i=1}^{N_r} \frac{[\Rb_{\Hbs_m}]_{i,j} \sigma_w^2 }{[\Rb_{\Hbs_m}]_{i,j} + \sigma_w^2}$, and ${\rm MSE}_j^{\rm c,AGE}$ is given by \eqref{eqn: mse central age algorithm}.
Moreover, we have ${\rm MSE}^{\rm AGE} \leq {\rm MSE}^{\rm FD}$. 
\end{prop}
%\begin{IEEEproof}
	{\emph Proof:} The proof is relegated to Appendix \ref{appd: proof of mse of age algorithm}. \hfill $\blacksquare$
%	\end{IEEEproof}

\begin{rem} \label{rem : aggregation stationary}
	It is interesting to see from Proposition \ref{prop : nmse aggregation} that, the AGE-based algorithm can realize flexible tradeoff between the estimation accuracy and inter-BBU communication cost by choosing different hard thresholding parameter $\eta$, and meanwhile include the centralized and FD algorithms as special cases. Specifically, by setting $\eta=0$, all the local information is uploaded to the central node, and thus the AGE-based algorithm becomes the centralized algorithm. While, for a sufficiently large $\eta$, no information is uploaded to the central node, and therefore each local node estimates its channel solely with its local received signal. Hence, the AGE-based algorithm degrades to the FD algorithm.
\end{rem}     

\begin{figure}[t]
	\centering
	\includegraphics[width=0.52\linewidth]{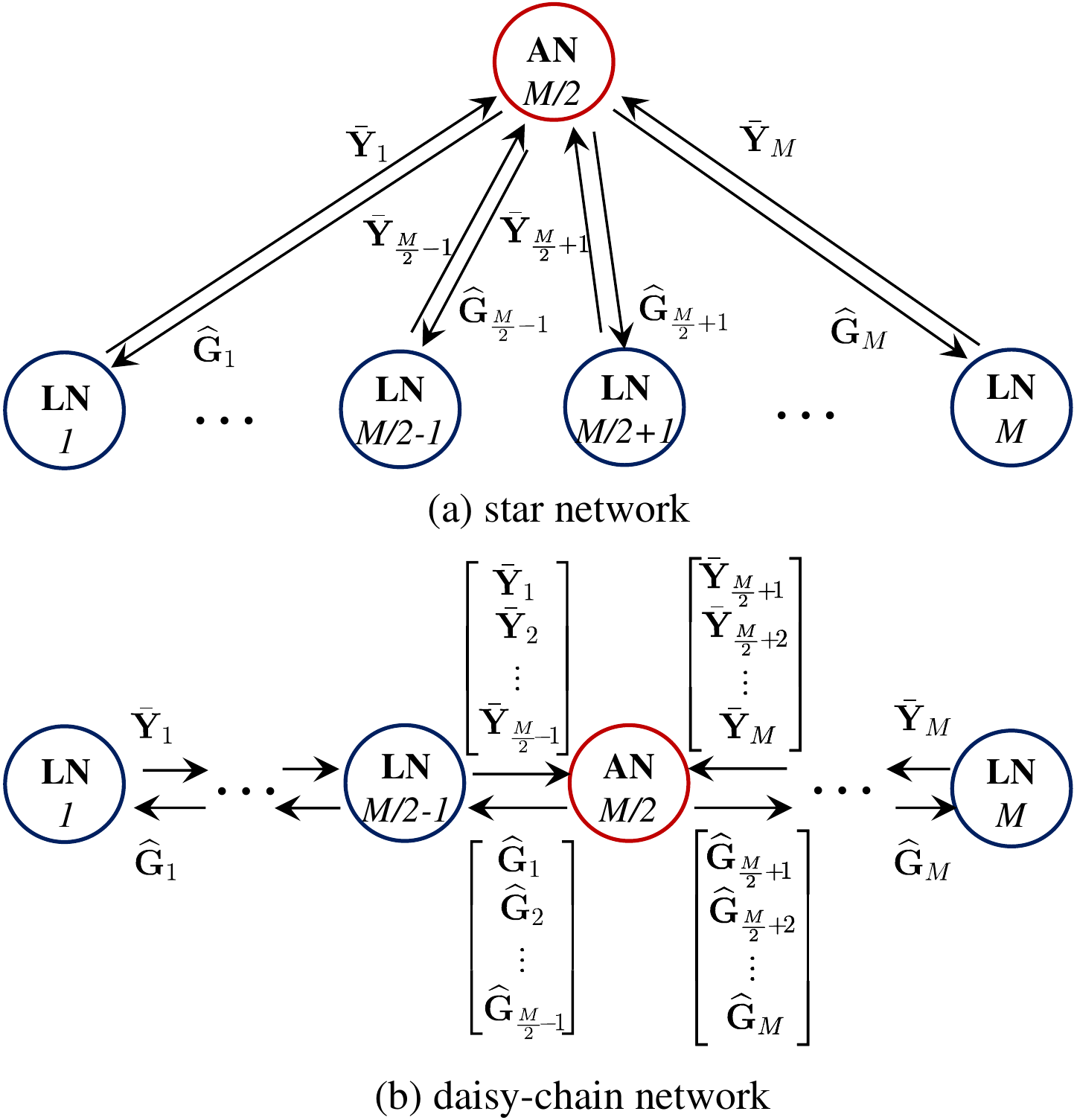}\\
	\caption{Illustration of the signal exchange processes of the AGE-based algorithm in (a) the star network and (b) the daisy-chain network, where ``AN'' denotes the aggregation node and ``LN'' represents a local node.} 
	\label{fig:information_exchange_stationary_dl_agg}
\end{figure}

\begin{rem}\label{rem: aggregation in chain}
	It is worth mentioning that the AGE-based algorithm can also be directly applied to the daisy-chain network. The only difference is how the signals are exchanged among nodes.
%	that the local nodes need to first accumulate the signals from the previous nodes together with its own and then forward to the next node in the daisy-chain network. 
	Without loss of generality, let node $\frac{M}{2}$ be the aggregation node in both the star and daisy-chain networks. Taking the uploading phase for example, in the star network, node $\frac{M}{2}-1$ uploads $\bar \Yb_{\frac{M}{2}-1}$ to the aggregation node; while, in the daisy-chain network, node $\frac{M}{2}-1$ needs to first accumulate the signals from the previous nodes together with its own  by $[\bar\Yb_1^\Hf,\bar\Yb_2^\Hf,...,\bar\Yb_{\frac{M}{2}\!-\!1}^\Hf\!]^\Hf$ and then upload them.
	For ease of understanding, the detailed signal exchange processes of AGE-based algorithms in the two networks are given in Fig. \ref{fig:information_exchange_stationary_dl_agg}.
\end{rem}

\subsection{EAG-Based Distributed Algorithm }
In this subsection, we present the EAG-based distributed algorithm, in which both the angle and delay-domain sparsity of the local channel are exploited and thus the inter-BBU communication cost is reduced compared to the AGE-based algorithm.
Let's consider the star network. 

{\bf \indent 1) Processing at local nodes: } Each local node first obtains its local channel estimate, $\widehat\Hbs_m,\forall m \in \Mset$, by \eqref{eqn: channel estimate local}. Then, hard windowing is applied to the local estimate and obtains
\begin{align}\label{eqn: EAG hard windowed H}
	\widehat\Gbs_m &\triangleq \Db_m \odot  \widehat\Hbs_m, \in \mathbb{C}^{N_r \times N_C},
\end{align} 
where $\Db_m$ is the hard windowing matrix. Notice that in \eqref{eqn: EAG hard windowed H} the local channel estimate is in its angle-and-delay domain, which is different from the AGE-based algorithm where only delay-domain sparsity of the local received signal is exploited. To utilize the angle-and-delay-domain sparsity for reducing the inter-BBU communication cost, $\Db_m$ can be designed as a column-and-row-sparse matrix.
Specifically, the $(i,j)$-th entry of $\Db_m$ is determined by
\begin{align}\label{eq:column sparsity cs dl}
	[\Db_m]_{i,j} = 
	\begin{cases}
		1, &{\rm if}~ 
			\begin{subarray}{l}
				\|[\widehat\Hbs_m ]_j\|^2 \geq \eta N_r \sigma_w^2, {\rm~ and} \\
				\|[\widehat\Hbs_m^\top]_i\|^2 \geq \eta N_C \sigma_w^2,
			\end{subarray}\\
		0, &{\rm otherwise},
	\end{cases}
\end{align}
With the hard-windowed channel estimate $\widehat \Gbs_m$, the residual channel estimate at local node $m$ can be written as
\begin{align}
	\widehat\Gbs_m^{(l)} = \widehat\Hbs_m - \widehat\Gbs_m \in \mathbb{C}^{N_r \times N_C}, m\in \mathcal{M},
\end{align}
%Then, the local nodes send $\widehat\Gbs_m,m\in \mathcal{M},$ to the central node. 
and let $\Jc_m^{(l)}$ be the index set of the non-zero entries of $\widehat{\Gbs}^{(l)}_{m}$. 

{\bf \indent 2) Processing at central node : }
%The central node first transforms $\widehat\Gbs_m,m\in \mathcal{M},$ to their antenna domain by
%\begin{align} \label{eqn: delay to angle at central node in EAG}
%		\widehat{\widetilde \Gb}_m = \Fb_{N_r} \widehat\Gbs_m,
%\end{align}
The local nodes upload the hard-windowed channel estimates $\widehat\Gbs_m$s', to the central node, which are then aggregated by
\begin{align}\label{eqn: aggregation central EAG}
	\widehat \Qbs = \sum\nolimits_{m=1}^M \Fb_m^\Hf \Fb_{N_r} \widehat\Gbs_m \in \mathbb{C}^{N_R \times N_C}.
\end{align}
%where $\widetilde \Fb_{N_r} = \Ib_M \otimes \Fb_{N_r} \in \Cs^{N_R \times N_R}$.
%
To further improve the estimation accuracy, the central node uses $\widehat \Qbs$ to refine the estimation by solving
\begin{align}
	\min_{\widetilde \Sbs} \E\left\{\|\widetilde\Sbs \odot \widehat\Qbs - \bar\Hbs\|_F^2\right\},
\end{align}
where $\bar \Hbs = \sum_{m=1}^{M}\Fb_m^\Hf \Fb_{N_r} (\Db_m\odot\Hbs_m)$. Analogous to \eqref{eqn: angle delay channel estimate of central diagonal mmse} and \eqref{eqn: channel estimate local}, the estimated channel is given by
\begin{align} \label{eqn: channel estimate EAG central node}
	\widehat\Bbs = \widetilde\Sbs \odot \widehat\Qbs,
\end{align}
where $[\widetilde\Sbs]_{i,j} = \frac{[\Rb_{\bar\Hbs}]_{i,j}}{[\Rb_{\widehat\Qbs}]_{i,j}}$ for $[\Rb_{\widehat\Qbs}]_{i,j} \neq 0$, $\Rb_{\bar\Hbs} = \E\{\bar\Hbs \odot \bar\Hbs^*\}$ and $\Rb_{\widehat\Qbs} = \E\{\widehat\Qbs \odot \widehat\Qbs^*\}$. 
The associated MSE of estimating the $(i,j)$-th entry is given by 
\begin{align} \label{eqn: mse EAG central node}
	{\rm MSE}_{i,j}^{\rm c,EAG} & = \E\left\{\left|[\widetilde\Sbs]_{i,j} [\widehat\Qbs]_{i,j} - [\bar\Hbs]_{i,j}\right|^2\right\} \notag\\
	&\overset{(a)}{=} [\widetilde\Sbs]_{i,j}^2[\Rb_{\widehat\Qbs}]_{i,j} -2[\widetilde\Sbs]_{i,j} [\Rb_{\bar\Hbs}]_{i,j} + [\Rb_{\bar\Hbs}]_{i,j}\notag\\
	&=\frac{[\Rb_{\bar\Hbs}]_{i,j}\left([\Rb_{\widehat\Qbs}]_{i,j}-[\Rb_{\bar\Hbs}]_{i,j}\right)}{[\Rb_{\widehat\Qbs}]_{i,j}},
\end{align}
where (a) is because of the assumption that $\Rb_{\widehat\Qbs\bar\Hbs} = \Rb_{\bar\Hbs}$. Due to the oversampling on $\widehat\Gbs_m$ by $\Fb_m^\Hf, \forall m\!\in\!\Mset,$ in \eqref{eqn: aggregation central EAG}, one can also conclude that ${\rm MSE}_{i,j}^{\rm c,EAG}$ is no larger than that incurred by the FD scheme as per Theorem \ref{theo : mse central and fd}.
%
%which is no larger than that incurred by FD estimation.

Then, the central node transforms $\widehat\Bbs$ to its antenna-delay domain by
	\begin{align}  \label{eqn: central antenna delay channel estimate}
		\widehat{\Bb} = \Fb_{N_R}\widehat\Bbs= [\widehat{\Bb}_1^\Hf,\widehat{\Bb}_2^\Hf,\ldots\widehat{\Bb}_M^\Hf]^\Hf \in \mathbb{C}^{N_R \times N_C},
	\end{align}
and $\widehat{\Bb}_m$s' are sent to the corresponding local nodes. 

{\bf \indent 3) Post-processing at local nodes : }
Then, $\widehat{\Bb}_m$s' are transformed to their angle-and-delay domains by
\begin{align} \label{eqn: local anle delay channel estimate at central node}
	\widehat\Bbs_m = \Fb_{N_r} \widehat{\Bb}_m \in \Cs^{N_r \times N_C}, \forall m\in\Mset.
\end{align} 

Similar to the AGE-based algorithm, each local node determines its local angle-and-delay-domain channel estimate based on $\widehat{\Bbs}_m$ and $\widehat{\Gbs}_m^{(l)}$.
%By letting $\Jc_m$ be the index set of the non-zero entries of $\widehat{\Bbs}_{m}$, we have 
%\begin{align}
%	\Jc_m \cup \Jc_m^{(l)} = \big\{(i,j)|&\forall i\in\{(m-1)N_r+1,...,mN_r\},\notag\\
%	& \forall j\in\{1,2,...,N_C\}\big\}, \forall m,
%\end{align}
%and {\small $\widetilde\Jc_m = \Jc_m \cap \Jc_m^{(l)} \neq \emptyset, \forall m$}.
%Analogous to the AGE-based algorithm, 
In particular, the $(i,j)$-th angle-and-delay-domain channel entry of the node $m$ is given by
	\begin{align}\label{eqn: anntenn delay domain channel of EAG algorithm}
	[\widehat{\Hbs}_m]_{i,j} \!=\! 
	\begin{cases}
	[\widehat{\Bbs}_m]_{i,j}, & {\rm for}~ (i,j) \in \bar\Jc_m,\\
	[\widehat{\Gbs}^{(l)}_m]_{i,j}, & {\rm for}~ (i,j) \in \bar\Jc_m^{(l)},\\
	\alpha [\widehat{\Bbs}_m]_{i,j} \!+\! (1\!-\!\alpha) [\widehat{\Gbs}^{(l)}_m]_{i,j},\!\!\!&
	{\rm otherwise},
	\end{cases}
	\end{align}
where 
%\begin{align*}
	{\small $\bar \Jc_m = \left\{(i,j)|(i,j) \in \Jc_m\setminus\Jc_m^{(l)}\right\},
		\bar \Jc_m^{(l)} = \left\{(i,j)|(i,j) \in \Jc_m^{(l)}\setminus\notin \Jc_m\right\}, \forall m \in \Mset$},
%\end{align*}
and $0\leq \alpha \leq 1$ can be chosen empirically.

Finally, the antenna-and-frequency domain channel estimate of each local node is given by
\begin{align}\label{eqn: EAG complete channel estimate}
	\widehat \Hb_m = \Fb_{N_r} \widehat{\Hbs}_{m} \Fb_{N_C} ,m\in \mathcal{M},
\end{align}
where $\widehat{\Hbs}_{m}$ is obtained by \eqref{eqn: anntenn delay domain channel of EAG algorithm}, and the complete channel estimate is $\widehat\Hb = [\widehat\Hb_1^\Hf,\widehat\Hb_2^\Hf,...,\widehat\Hb_M^\Hf]^\Hf$. The details of the EAG-based distributed CE algorithm are summarized in Algorithm \ref{alg:EAG-based CE}. 
\begin{algorithm}[!tb] \small 
	\caption{EAG-based algorithm for distributed CE}
	\begin{algorithmic}[1]
		\REQUIRE Received antenna-and-frequency-domain signal $\Yb$, noise power $\sigma_w^2$, number of clusters $M$, and threshold parameter $\eta$. \\
		{\indent\bf 1) Processing at local nodes:}
		\STATE {Each node first estimates the local channel and then obtains the hard-windowed {\smaller $\widehat\Gbs_m$} by \eqref{eqn: EAG hard windowed H}.}\\
		{\bf 2) Processing at central node:}\vspace{0mm}
		\STATE {The central node collects {\smaller $\widehat{\Gbs}_m$}s' and aggregates them by \eqref{eqn: aggregation central EAG}.}
		\STATE {The central node re-estimates the channel by \eqref{eqn: channel estimate EAG central node}}
		\STATE {The central node transforms the centrally estimated channel to local angle-and-delay domain channel by \eqref{eqn: central antenna delay channel estimate} and \eqref{eqn: local anle delay channel estimate at central node}}\\
		{\bf 3) Post-processing at local nodes:}
		\STATE {Each node obtains its angle-and-delay-domain channel by \eqref{eqn: anntenn delay domain channel of EAG algorithm}.}
		\STATE {Finally, the local antenna-and-frequency-domain channel estimate is obtained by \eqref{eqn: EAG complete channel estimate}}
		\ENSURE {The estimated local channels: $\widehat \Hb_m,m\in \mathcal{M}.$}
	\end{algorithmic}\label{alg:EAG-based CE} 
\end{algorithm}%\vspace{-3mm}
The EAG-based algorithm is also applicable to the daisy-chain network. 

\begin{prop} \label{prop : mse EAG algorithm}
	The MSE of the EAG-based distributed CE algorithm is upper-bounded by
	\begin{align}\label{eqn: mse EAG algorithm}
			{\rm MSE}^{\rm EAG} \leq& \sum_{(i,j)\in \bar\Jc_m} {\rm MSE}_{i,j}^{\rm c,EAG} + \sum_{(i,j)\in \bar\Jc_m^{(l)}} {\rm MSE}_{i,j}^{\rm FD} \notag\\
			&+ \sum_{(i,j)\in \Jc_m \cap \Jc_m^{(l)}} \left(\alpha {\rm MSE}_{i,j}^{\rm c,EAG}+ (1-\alpha){\rm MSE}_{i,j}^{\rm FD}\right),
	\end{align}
where  ${\rm MSE}_{i,j}^{\rm FD} \triangleq \sum_{m=1}^M \frac{[\Rb_{\Hbs_m}]_{i,j} \sigma_w^2 }{[\Rb_{\Hbs_m}]_{i,j} + \sigma_w^2}$ and ${\rm MSE}_{i,j}^{\rm c,EAG}$ is given in \eqref{eqn: mse EAG central node}.  Besides, we have ${\rm MSE}^{\rm EAG} \leq {\rm MSE}^{\rm FD}$. 
\end{prop}
%\begin{IEEEproof}
	{\emph Proof:} The proof is similar to that of Proposition \ref{prop : nmse aggregation}. We omit it due to the limited space. \hfill $\blacksquare$
%\end{IEEEproof}

\begin{rem}
	Notice that, the EAG-based algorithm can exploit both the angle- and delay-domain sparsities.
	Hence, it can achieve a similar performance as the AGE-based algorithm but with an even smaller inter-BBU communication cost. As a tradeoff, the computation complexity of the EAG-based algorithm is slightly higher than the AGE-based algorithm due to the refined estimation at the aggregation node.
\end{rem}

\section{Communication and Complexity Analysis} \label{sec: communication and complexity}
In this section, the fronthaul/inter-BBU communication costs and computation complexities of the baseline schemes and the proposed algorithms are analyzed.

\subsection{Fronthaul/Inter-BBU Communication Cost} \label{subsec : communication cost}
In this subsection, we analyze the inter-BBU communication costs of the baseline schemes and the proposed distributed algorithms. The inter-BBU communication cost is measured by the number of real values exchanged among nodes during the whole procedures of the algorithms.

\subsubsection{Fronthaul communication cost of the centralized algorithm} 
%Let's first consider the centralized algorithm.
In the star network, any local node $m \in \Mset$ can be set as the central node. Without loss of generality, let node $M$ as the central node.
Then, the other $M-1$ local nodes upload $\Yb_m \in \Cs^{N_r \times N_C}, m\in\Mset\setminus M$, to node $M$ for centralized CE and then the estimated channel $\widehat{\Hb}_m\in \Cs^{N_r \times N_C}, m\in\Mset\setminus M$, will be returned to the corresponding local nodes. 
Following the above process, the total number of exchanged real values is {\small$4(M-1)N_RN_C/M$}. 
While, in the daisy-chain network, we let node {\small$M/2$} be the destination node.
Since the information can only be exchanged between the neighboring nodes,
the information of node $m$ should be forwarded {\small $\left|M/2 - m\right|$} times to reach node {\small$M/2$}. As a result, the total number of real values exchanged in the daisy-chain network for both uploading and downloading is {\small$MN_RN_C$}.
Summarily, the total numbers of exchanged real values for the centralized scheme in star and daisy-chain networks are respectively given by
\begin{subequations} \label{eq:dl agg total bits for centralized scheme}
	\begin{align}
		N_{\rm c}^{\rm star}  &= \frac{4(M-1)N_RN_C}{M},\\
		N_{\rm c}^{\rm chain} &= 4\sum_{m=1}^{M} \frac{\left|M/2 - m\right|N_RN_C}{M} = MN_RN_C,
	\end{align}
\end{subequations}% 	
where $|x|$ denotes the absolute value of $x$.

\begin{table*}[!t]
	\centering\smaller[3]
	\caption{$R_c^{t}$ ($R_c^{dn}$) of the AGE/EAG-based algorithms.}\renewcommand\arraystretch{1.8}
	\begin{tabular}{l c c c c c}   
		\hline
		& $M=2$   & $M=4$   & $M=8$   & $M=16$  \\
		\hline
		$\bar N_{C} = \frac{N_{C}}{100} $ & $0.901/0.903$ ($0.0022/0.0028$)  & $0.851/0.853$ ($0.0024/0.0029$) & $0.827/0.828$ ($0.0024/0.0030$) & $0.814/0.815$ ($0.0025/0.0031$) \\
		%		\hline
		$\bar N_{C} = \frac{N_C}{20} $ & $0.905/0.915$ ($0.0111/0.0137$)  & $0.857/0.863$ ($0.0117/0.0145$)  & $0.834/0.838$ ($0.0120/0.0149$)  & $0.822/0.825$ ($0.0122/0.0152$) \\
		%		\hline
		$\bar N_{C} = \frac{N_{C}}{10} $ & $0.910/0.930$ ($0.0220/0.0269$)  & $0.867/0.877$ ($0.0231/0.0285$)  & $0.842/0.851$ ($0.0238/0.0294$)  & $0.831/0.838$ ($0.0241/0.0299$) \\
		%		\hline
		$\bar N_{C} = \frac{N_C}{5}$ & $0.920/0.960$ ($0.0435/0.0521$)  & $0.880/0.905$ ($0.0445/0.0553$)  & $0.860/0.877$ ($0.0455/0.0570$)& $0.850/0.864$ ($0.0471/0.0579$) \\
		\hline
	\end{tabular}\label{tab:computatiuonal complexity ratio aggregation scheme}
\end{table*}

\subsubsection{Inter-BBU communication cost of the proposed distributed algorithms}
Notice that, to reconstruct a sparse signal at the destination node, each local node should send the values and corresponding indices of the nonzero channel entries after hard windowing. 
Thanks to the special structure of the sparse signal (column sparse in the AGE-based algorithm, row-and-column sparse in the EAG-based algorithm), one only need to send the values and their corresponding indices of the nonzero columns and rows to the destination node. 
Let's first consider the AGE-based algorithm. Suppose that the number of nonzero columns after hard windowing for uploading and downloading are $\bar N_{C_m}^{\rm ul}$ and $\bar N_{C_m}^{\rm dl},m\in\Mset$, respectively. Then, the total number of real values exchanged in the star and daisy-chain networks are respectively given by
\begin{subequations}\label{eq:dl agg total bits for agg scheme}
	\begin{align}
		N_{\rm AGE}^{\rm star} &= \underbrace{2N_r \sum_{m=1}^{M-1} \left(\bar N_{C_m}^{\rm ul} \!+\! \bar N_{C_m}^{\rm dl}\right)}_{\rm nonzero~channel~entries}+ \underbrace{\sum_{m=1}^{M-1} \left(\bar N_{C_m}^{\rm ul} \!+\! \bar N_{C_m}^{\rm dl}\right)}_{\rm indices~of~nonzero~columns} \notag\\
		&=(2N_r+1) \sum\nolimits_{m=1}^{M-1} \left(\bar N_{C_m}^{\rm ul} \!+\! \bar N_{C_m}^{\rm dl}\right),\\
		N_{\rm AGE}^{\rm chain} &=  \underbrace{2N_r \sum\nolimits_{m=1}^M \left|M/2-m\right|\left(\bar N_{C_m}^{\rm ul} + \bar N_{C_m}^{\rm dl}\right)}_{\rm nonzero~channel~entries} + \underbrace{\sum\nolimits_{m=1}^M \left|M/2-m\right|\left(\bar N_{C_m}^{\rm ul} + \bar N_{C_m}^{\rm dl}\right)}_{\rm indices~of~nonzero~columns} \notag\\
		&= (2N_r + 1)\sum\nolimits_{m=1}^M \left|M/2-m\right|\left(\bar N_{C_m}^{\rm ul} + \bar N_{C_m}^{\rm dl}\right).
	\end{align}
\end{subequations}%
For a more clear comparison, we define the communication cost of an algorithm as the ratio of the number of exchanged real values of the algorithm to that of the centralized algorithm. Combining \eqref{eq:dl agg total bits for centralized scheme} and \eqref{eq:dl agg total bits for agg scheme}, the communication costs of the AGE-based algorithm in star and daisy-chain networks are respectively given by 
\begin{subequations}\label{eq:dl age communication cost}
	\begin{align}
	{\rm C_{\rm AGE}^{\rm star}} &= \frac{N_{\rm AGE}^{\rm star}}{N_c^{\rm star}} =  \frac{(2N_r + 1) \sum_{m=1}^{M-1} \left(\bar N_{C_m}^{\rm ul} + \bar N_{C_m}^{\rm dl}\right)}{4N_RN_C(M-1)/M},\\
	{\rm C_{\rm AGE}^{\rm chain}} &= \frac{N_{\rm AGE}^{\rm chain}}{N_c^{\rm chain}} = \frac{ (2N_r + 1)\sum_{m=1}^M \left|\frac{M}{2}-m\right|\left(\bar N_{C_m}^{\rm ul} + \bar N_{C_m}^{\rm dl}\right)}{MN_RN_C}.
	\end{align}
\end{subequations}%
While, for the EAG-based algorithm, the communication costs in the star and daisy-chain network are respectively given by 
\begin{subequations}\label{eq:dl cs communication cost}
	\begin{align}
		{\rm C_{\rm EAG}^{\rm star}} =& \frac{\sum\limits_{m=1}^{M-1} \left(2\bar N_{r_m}^{\rm ul}\bar N_{C_m}^{\rm ul}+\bar N_{C_m}^{\rm ul}+\bar N_{r_m}^{\rm ul} \right) + 
			(2N_r +1)\sum\limits_{m=1}^{M-1} \!\bar N_{C_m}^{\rm dl}}{4N_RN_C(M-1)/M},\\
		{\rm C_{\rm EAG}^{\rm chain}} =& \Big(\sum_{m=1}^{M} \left|M/2-m\right| \left(2\bar N_{r_m}^{\rm ul}\bar N_{C_m}^{\rm ul} + \bar N_{C_m}^{\rm ul} +\bar N_{r_m}^{\rm ul} \right)  \notag\\
			&+ (2N_r+1)\sum_{m=1}^{M} \left|M/2-m\right| \bar N_{C_m}^{\rm dl}\Big)/(MN_RN_C),
	\end{align}
\end{subequations}%
where $\bar N_{r_m}^{\rm ul}$ and $\bar N_{C_m}^{\rm ul}$ denote the number of preserved rows and columns of local node $m$ after hard windowing, while $\bar N_{C_m}^{\rm dl}$ denotes the preserved columns for downloading. Notice that in the downloading phase, the central node sends the antenna-delay domain channel, i.e., {\small$\widehat{\Bb}_m$s'}, to the local node and there is no row sparsity in {\small$\widehat{\Bb}_m$s'}.
In Section \ref{sec: numerical result}, the NMSE performance versus different inter-BBU communication costs of the proposed algorithms will be evaluated in detail.

\begin{table}[t]
	\centering \smaller[1]
	\caption{Computation complexity of the baseline schemes and the proposed algorithms.} \renewcommand\arraystretch{1.6}
	\begin{tabular}{ c| c }   
		\hline 
		& Computation Complexity\\
		\hline
		C-DMMSE  &$8N_R^2N_C + 8N_RN_C^2 + 2N_RN_C$\\
		\hline
		FD-DMMSE &   $8N_rN_RN_C + 8N_RN_C^2 + 2N_RN_C$\\
		\hline
		\multirow{1}*{AGE } & \multirow{2}*{$\underbrace{4N_R^2 \bar N_C}_{\eqref{eqn: aggregation central windowing}} + \underbrace{2 N_R\bar N_C}_{\eqref{eq: dl central estimation}} + \underbrace{4N_R^2 \bar N_C}_{\eqref{eq: dl central angle to antenna}}$}\\
		central node& \\
		\hline
		AGE  & \multirow{2}*{$\underbrace{4N_RN_C^2}_{\eqref{eqn: aggregation local windowing}}\!+\! \underbrace{4N_RN_C^2}_{\eqref{eq: dl agg complete channel estiamte}}\!+\!\underbrace{(8N_r N_R \!+\! 2N_R) (N_C \!-\! \bar N_C)}_{{\rm estimation~ of ~ } \widehat \Gb_m^{(l)}, \forall m\in \Mset}$}\\
		local nodes& \\
		\hline	
		EAG  & \multirow{2}*{$\underbrace{4N_R \bar N_r \bar N_C + 4N_R^2\bar N_C}_{\eqref{eqn: aggregation central EAG}} + \underbrace{2N_R \bar N_C}_{\eqref{eqn: channel estimate EAG central node}}$ + $\underbrace{4N_R^2\bar N_C}_{\eqref{eqn: central antenna delay channel estimate}}$}\\
		central node& \\
		\hline
		EAG   & \multirow{2}*{{$\underbrace{4N_rN_RN_C + 4N_RN_C^2 + 2N_RN_C}_{\eqref{eqn: EAG hard windowed H}}$ $+$ $\underbrace{4N_r N_R\bar N_C}_{\eqref{eqn: local anle delay channel estimate at central node}} + \underbrace{4N_r N_R N_C + 4N_R N_C^2}_{\eqref{eqn: EAG complete channel estimate}}$}} \\
		local nodes& \\
		\hline	
	\end{tabular}\label{tab:computatiuonal complexity age scheme}
\end{table}

\subsection{Computation Complexity}
In this subsection, we compare the computation complexity of the proposed algorithms with the baseline schemes. The computation complexity is measured by the number of real-valued multiplications.
Take the centralized algorithm for example, the computations occurs in the following three steps:
\begin{itemize}
	\item {\emph{IDFT Transformation:}} The antenna-and-frequency-domain received signal is first transformed into the angle-and-delay-domain by $\Ybs = \Fb_{N_R}^\Hf \Yb \Fb_{N_C}^\Hf$. Thus, the associated operations take a computation of $4N_R^2N_C + 4N_RN_C^2$.
	\item {\emph{CE:}} The computation complexity comes from the soft windowing operation to estimate the angle-and-delay-domain channel by $\widehat{\Hbs} = \Sbs \odot \Ybs$. Thus, the corresponding computation complexity is $2N_R N_C$.
	\item {\emph{DFT Transformation:}} The estimated angle-and-delay-domain channel is transformed to antenna-and-frequency-domain by $\widehat\Hb = \Fb_{N_R} \widehat\Hbs \Fb_{N_C}$, which takes the same computation complexity as the IDFT transformation and is given by $4N_R^2N_C + 4N_RN_C^2$.
\end{itemize}
Summarily, the total computation complexity of the centralized scheme is $8N_R^2N_C + 8N_RN_C^2 + 2N_RN_C$. 
The computation complexities of the FD algorithm and the proposed distributed algorithms can be calculated in the same way, which are summarized in Table \ref{tab:computatiuonal complexity age scheme}. 
We note that the computation complexities of proposed algorithms in Table \ref{tab:computatiuonal complexity ratio aggregation scheme} are obtained with the assumption that the nodes have same sparsity pattern.

To give a more clear comparison, we denote $R_c^{t}$ as the ratio of the total computation complexity of the proposed algorithms to that of the centralized scheme. The values of $R_c^{t}$ under different parameter settings are summarized in
Table \ref{tab:computatiuonal complexity ratio aggregation scheme}. We use $\bar N_C$ to denote the number of selected columns that are exchanged between aggregation node and local node in both the uploading and downloading phases. Besides, for the EAG-based algorithm, we assume that half of the rows are selected to send to the aggregation node.
As it can be seen from Table \ref{tab:computatiuonal complexity ratio aggregation scheme} that the computational complexities of proposed algorithms are smaller than that of the centralized scheme. 
Besides, the computational complexities of the proposed algorithms are robust to the number antenna clusters and hard windowing.
Remind that, in both the star network and daisy-chain network, there is a destination node to perform centralized CE. It is also interesting to see the ratio, denoted as $R_c^{dn}$, of the computation complexity at the destination node to the total computation complexity. One can see that the computation complexity at the destination node is only a small portion of that in total.
As a result, the computations are (approximately) evenly distributed across multiple nodes, and the computation complexity of each node is greatly reduced compared to the centralized algorithm. 
It can also be observed that the computation complexity of the EAG-based algorithm is slightly higher than the AGE-based algorithm both in the destination node and in total.
%
%\begin{figure*}[ht]
%	\centering
%	\subfigure[$\SNR = -20$dB]{\includegraphics[width=0.32\linewidth]{mse_agg_lowSNR_star} \label{fig:mse-gap-agg-lowsnr}}
%	\subfigure[$\SNR = 0$dB]{\includegraphics[width=0.32\linewidth]{mse_agg_midSNR_star} \label{fig:mse-gap-agg-midsnr}}
%	\subfigure[$\SNR = 20$dB]{\includegraphics[width=0.32\linewidth]{mse_agg_highSNR_star} \label{fig:mse-gap-agg-highsnr}}\vspace{-3mm}
%	\caption{ NMSE gap versus communication cost of AGE-based algorithm in the star network.}\vspace{-5mm}
%	\label{fig:mse communication downlink agg star}
%\end{figure*}
%\begin{figure*}[t!]
%	\centering
%	\subfigure[$\SNR = -20$dB]{\includegraphics[width=0.32\linewidth]{mse_agg_lowSNR_chain} \label{fig:mse-gap-agg-lowsnr}}
%	\subfigure[$\SNR = 0$dB]{\includegraphics[width=0.32\linewidth]{mse_agg_midSNR_chain} \label{fig:mse-gap-agg-midsnr}}
%	\subfigure[$\SNR = 20$dB]{\includegraphics[width=0.32\linewidth]{mse_agg_highSNR_chain} \label{fig:mse-gap-agg-highsnr}}\vspace{-3mm}
%	\caption{ NMSE gap versus communication cost of AGE-based algorithm in the chain network.}\vspace{-5mm}
%	\label{fig:mse communication downlink agg chain}
%\end{figure*}

\section{Numerical Results} \label{sec: numerical result}
In this section, we evaluate the performances of the proposed AGE and EAG-based distributed CE algorithms. In the simulations, the channel is generated by the ``3GPP-38.901-UMa-NLOS'' model in ``{\tt QuaDRiGa}'' \cite{quadriga} and the key parameters are summarized in Table \ref{table : simulation parameters}. The channel power profile, i.e., $\Rb_{\Hbs}$, is assumed to be known. In particular, it is approximated by
$\Rb_{\Hbs} = \frac{1}{L}\sum_{\ell=1}^L \Hbs_{\ell} \odot \Hbs_{\ell}^\Hf,$ 
where $\Hbs_{\ell} \in \Cs^{N_R\times N_C}$ represents the $\ell$-th channel realization and $\Rb_{\Hbs_m}$s' are obtained in the same way.
We set $L=10$ and $\alpha = 0.5$ (in \eqref{eqn: anntenn delay domain channel of AGE algorithm} and \eqref{eqn: anntenn delay domain channel of EAG algorithm}) in the following simulations.
%The CE accuracy is measured by the normalized mean square error (NMSE) of the estimated channel, which is given by ${\rm NMSE} = \|\Hb - \widehat \Hb\|_F^2/\|\Hb\|_F^2$.

\subsection{Performance of the AGE-Based Algorithm} 
\begin{figure}[t]
	\centering
	{\includegraphics[width=0.6\linewidth]{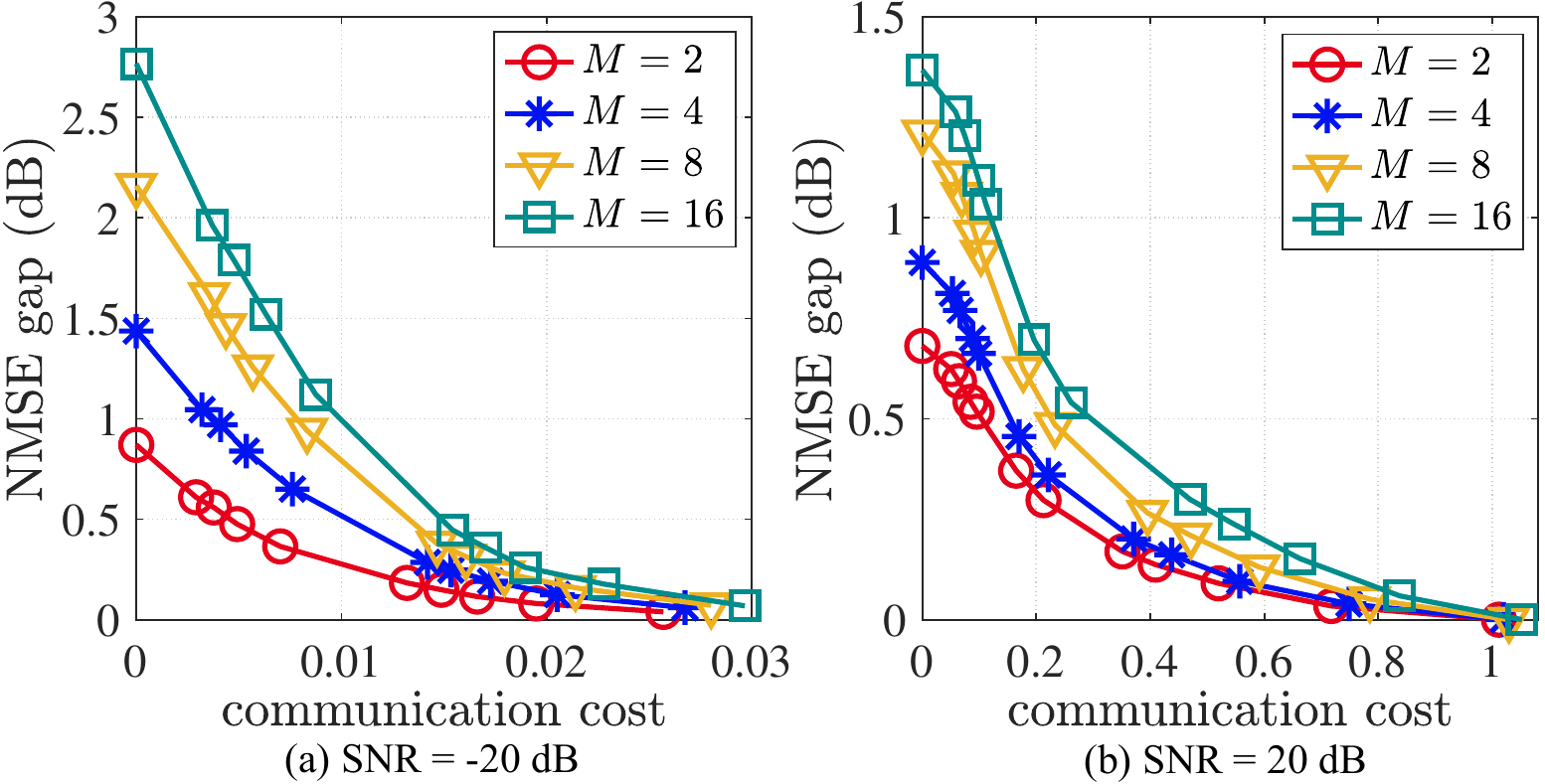}}
	\caption{NMSE gap versus inter-BBU communication cost of AGE-based algorithm in the star network.}
	\label{fig: nmse age star network} 
\end{figure}

\begin{figure}[t]
	\centering
	{\includegraphics[width=0.6\linewidth]{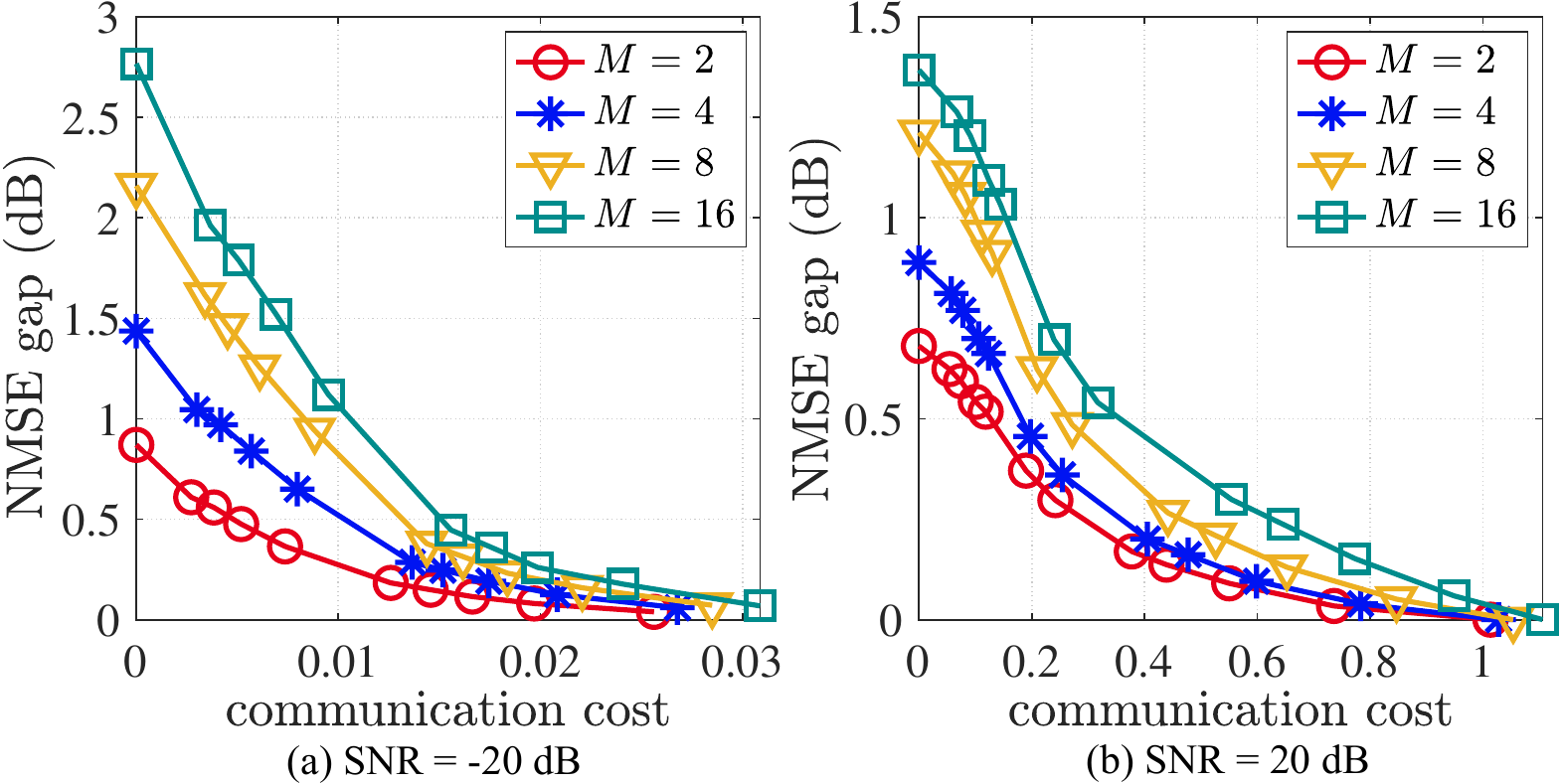}}
	\caption{NMSE gap versus inter-BBU communication cost of AGE-based algorithm in the daisy-chain network.}
	\label{fig: nmse age chain network}
\end{figure}

We first evaluate the performance of the AGE-based algorithm.
The NMSE performances versus inter-BBU communication costs of the AGE-based algorithm in star and daisy-chain networks are displayed in Fig. \ref{fig: nmse age star network} and Fig. \ref{fig: nmse age chain network}, respectively. 
The $y$-axis represents the NMSE gap between the AGE-based algorithm and the centralized algorithm.
The $x$-axis is the inter-BBU communication cost computed by \eqref{eq:dl age communication cost} by setting different hard windowing parameter $\eta$. 
Specifically, for $\SNR = -20$dB and $\SNR = 20$dB, $\eta$ are set as and
$[10^8,0.9,0.7,0.5,0.3,0.1,0.08,0.06,0.04,0.02]$ and $ [10^8,30,15,7,5,1,0.5,0.17,0.13,0.09,0.05,0.01]$, respectively.
In figures  \ref{fig: nmse age star network} and \ref{fig: nmse age chain network}, the cases of zero communication cost correspond to $\eta = 10^8$. In these cases, no local information is uploaded to the central node and the AGE-based algorithm degrades to the FD scheme. 
\begin{table}[t]
	\centering \smaller[1]
	\caption{Summary of the channel generating parameters}
	\label{table : simulation parameters}  \renewcommand\arraystretch{1.0}
	\begin{tabular}{c|c} %{cp{5.8cm}}  
		\hline%\hline
		parameter &value\\  
		\hline
		Number of antennas at BS& $256$\\
		Polarization of the antennas & $45^\circ/-45^\circ$ dual-polarized\\
		Antenna spacing & half wavelength\\
		Number of antennas at user& $1$\\
		Bandwidth& $100$MHz\\
		Number of subcarriers& $1024$\\
		Center frequency & $3.5$GHz\\
		3D-position of BS  &  $[0, 0, 25]$ (in meter) \\
		3D-position of user & $[200, 40, 1.5]$ (in meter) \\
		\hline %\hline
	\end{tabular}
\end{table}%
From figures \ref{fig: nmse age star network} and \ref{fig: nmse age chain network}, one can have the following interesting observations:
\begin{itemize}
	\item The AGE-based algorithm can realize flexible tradeoff between NMSE performance and inter-BBU communication cost.
	\item The NMSE performance {improves at the expense of the} inter-BBU communication cost as more channel entries are estimated centrally.
	\item The AGE-based algorithm can perform as well as the centralized scheme with significantly reduced inter-BBU communication cost. For example, in the low SNR regime (SNR = $-20$dB), with only an inter-BBU communication cost of $3\%$, the NMSE gap is smaller than $0.1$dB. 
	\item The required inter-BBU communication cost increases as the increase of SNR. This is due to the fact that, in the high SNR cases, more channel entries have comparable channel powers to the noise power. Consequently, more channel entries should be estimated centrally to approach the performance of the centralized scheme. 
	\item The AGE-based algorithm works well in both the daisy-chain and star networks.
\end{itemize}

\begin{figure}[t]
	\centering
	{\includegraphics[width=0.6\linewidth]{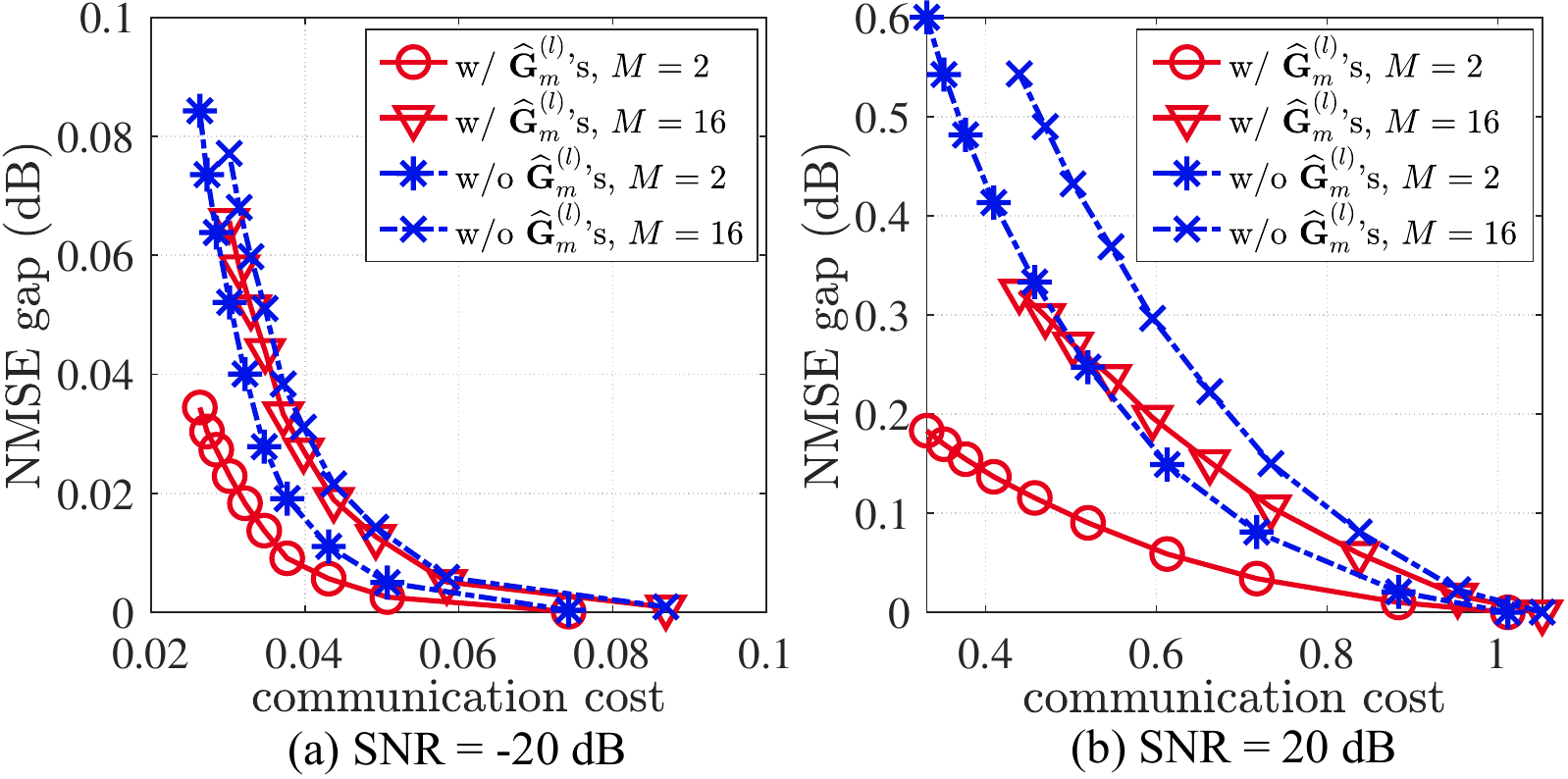}}
	\caption{ NMSE comparison of the AGE-based algorithm for the case with or without local estimation in the star network.} 
	\label{fig: mse with or without local information}
\end{figure}

In Fig. \ref{fig: mse with or without local information}, we verify whether the local estimate   $\widehat \Gb_m^{(l)}$s' in \eqref{eqn: anntenn delay domain channel of AGE algorithm} is really helpful to improve the estimation performance.  One can see from this figure, simply ignoring $\widehat \Gb_m^{(l)}$s' can greatly degrade the estimation accuracy, especially for the high SNRs.

\subsection{Comparison of the AGE and EAG-Based Algorithms}
\begin{figure}[t]
	\centering
	{\includegraphics[width=0.6\linewidth]{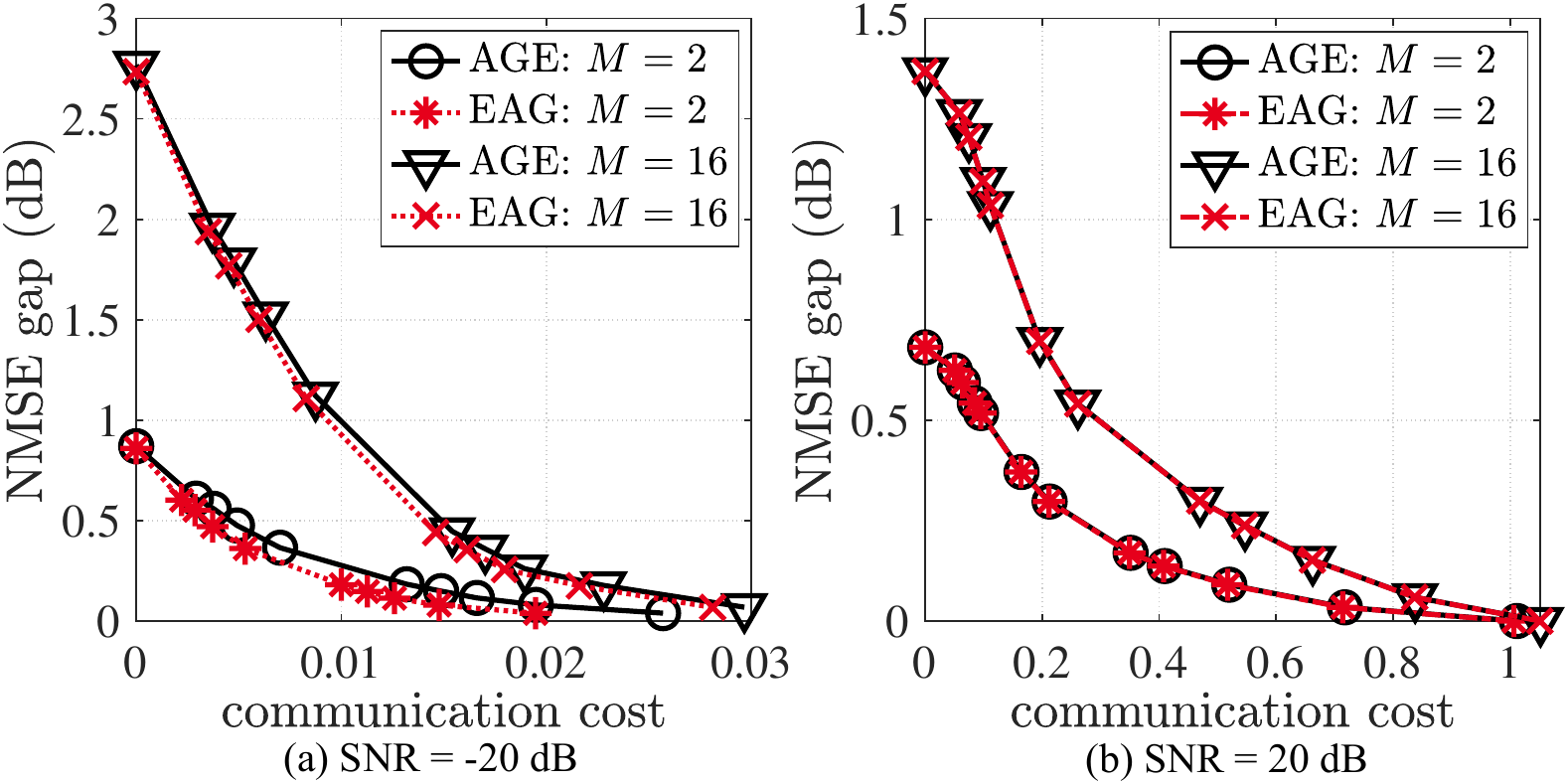}}
	\caption{NMSE performance comparison of AGE and EAG-based algorithms in the star network.}
	\label{fig: nmse EAG star network}
\end{figure}

\begin{figure}[t]
	\centering
	{\includegraphics[width=0.6\linewidth]{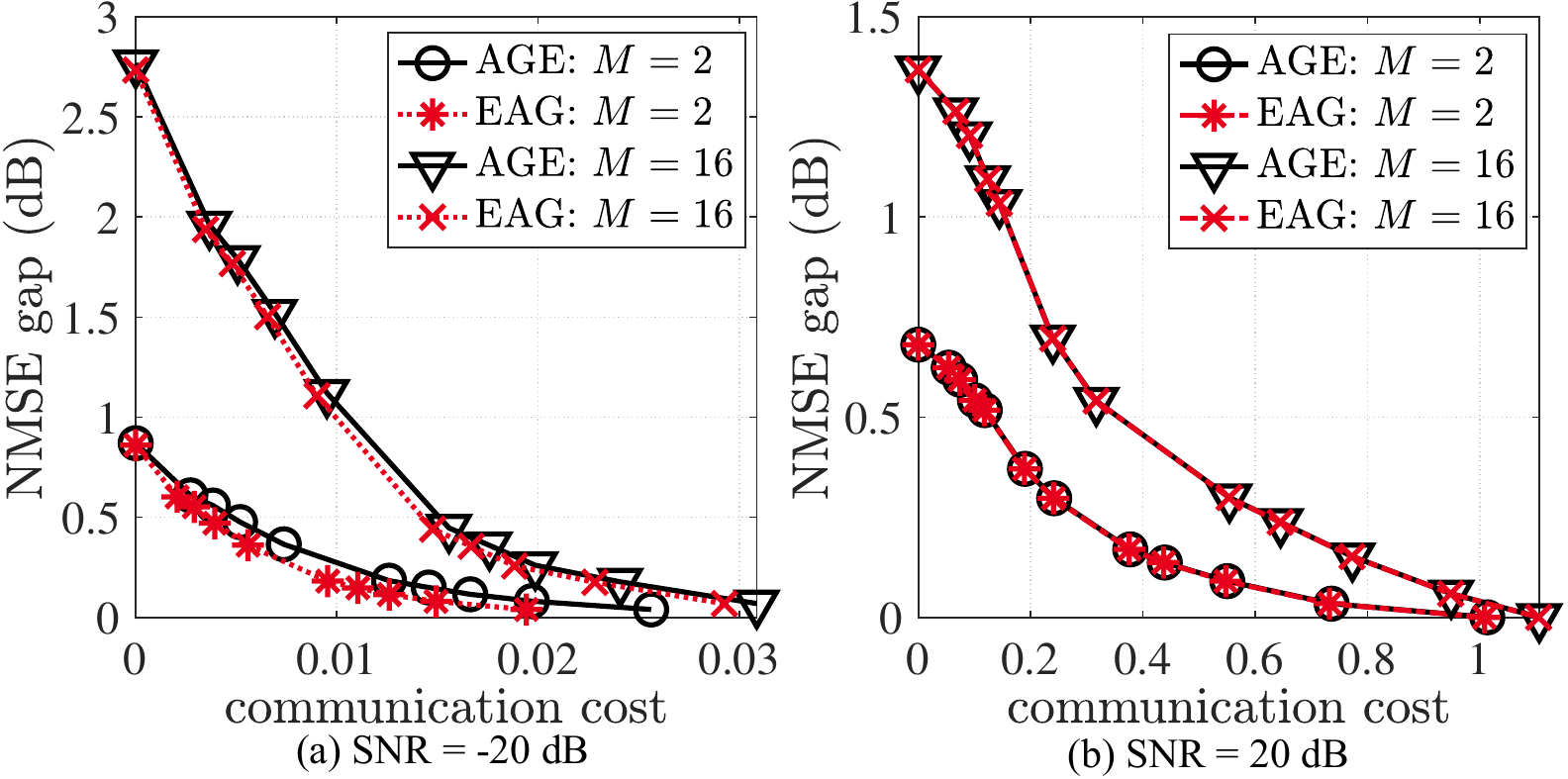}}
	\caption{NMSE performance comparison of AGE and EAG-based algorithms in the daisy-chain network.}
	\label{fig: nmse EAG chain network}
\end{figure}

The performance comparisons between the AGE and EAG-based algorithms are shown in Fig. \ref{fig: nmse EAG star network} and \ref{fig: nmse EAG chain network}. 
One can see that the EAG-based algorithm outperforms the AGE-based algorithm in the low SNR cases by benefiting the angular domain sparsity of the channel. 
Moreover, the performance gain is reduced as the increase of number of clusters because more antennas in a cluster will bring a higher resolution in the angular domain.
While, for the case of high SNR, the two algorithms perform closely since the angular domain sparsity becomes less noticeable as the increase of SNR. 

\section{Conclusions} \label{sec: conclusion}
In this paper, we have investigated the distributed CE algorithm design in the massive MIMO system under the DBP architecture. 
The low-complexity DMMSE estimator has been used as the baseline.
Firstly, we have theoretically proved that, by the DMMSE estimator, it is more accurate to estimate the channel from the angle-and-delay domain instead of the antenna-and-frequency domain by benefiting the power concentration phenomena of the angle-and-delay-domain channels.
Then, we have also proved that the centralized scheme strictly outperforms the FD scheme based on the DMMSE estimator. 
Then, by exploring the decomposable structure of the centralized scheme and the sparsities of the  channel in the angle and delay domains, two low-complexity distributed CE algorithms, i.e., the AGE-based algorithm and EAG-based algorithm, have been proposed. 
Extensive numerical simulations have shown that the proposed algorithms can perform as well as the centralized scheme but with a quite small communication cost in both the high and low SNRs. Besides, the two algorithms have also been shown to have lower computation complexity than the centralized scheme.

% to validate the advantages of proposed algorithms from the perspectives of the estimation accuracy, inter-BBU communication cost and computation complexity.
%can approach the centralized scheme both in low and high SNR regions with quite limited communication costs. 


\begin{appendices}
	\section*{\huge Appendices}
	\section{Proof of Theorem \ref{theo: mses of diagonal mmse algorithm}}\label{appd : proof of theorem I}
		We first present the following lemma.
		\begin{lem} \label{lem : concavity}
			Let $f(x)$ be a monotonically increasing and strictly concave function. Then for two sets of positive real numbers $\{a_k\}_{k=1}^K$ and $\{b_k\}_{k=1}^K$ which are arranged in the ascending order, i.e., $a_k \leq a_{k+1}$ and $b_k \leq b_{k+1}$ for $1 \leq k \leq K-1$, we have 
			\begin{align}
				\sum_{k=1}^K f(a_k) \leq \sum_{k=1}^K f(b_k),
			\end{align} 
			if $a_1 \leq b_1$, ${a_K} \geq {b_K}$, and $\sum_{k=1}^K a_k = \sum_{k=1}^K b_k$ are satisfied.
		\end{lem}
	
		The proof of Lemma \ref{lem : concavity} is relegated to Appendix \ref{appd : proof of lemma 2}.
		
		Next, we will show how to exploit Lemma \ref{lem : concavity} to prove Theorem \ref{theo: mses of diagonal mmse algorithm}. To this end, we first remind that
		\begin{align} \label{eqn: fact on channel power}
			\sum_{i=1}^{N_R}\sum_{j=1}^{N_C} [\Rb_{\Hbs}]_{i,j} = \sum_{i=1}^{N_R}\sum_{j=1}^{N_C} [\Rb_{\Hb}]_{i,j},
		\end{align}
		Then, recall ${\rm MSE}^{\rm c}$ and ${\rm MSE}^{\rm af}$ in \eqref{eqn: mse angle delay domain channel central} and \eqref{eqn: mse antenna frequency domain channel central}, respectively, and define
		\begin{align}
			f(x) = \frac{x\sigma_w^2 }{x+\sigma_w^2}.
		\end{align}
		It is not difficult to validate that $f(x)$ is strictly increasing and concave.
		Then, we can denote ${\rm MSE}^{\rm c}$ and ${\rm MSE}^{\rm af}$ by
		\begin{subequations}
			\begin{align}
				{\rm MSE}^{\rm c} &= \sum_{i=1}^{N_R}\sum_{j=1}^{N_C} f([\Rb_{\Hbs}]_{i,j}) \\
				{\rm MSE}^{\rm af} &= \sum_{i=1}^{N_R}\sum_{j=1}^{N_C} f([\Rb_{\Hb}]_{i,j})
			\end{align}
		\end{subequations}
		Then, as per \eqref{eqn: fact on channel power}, Assumption \ref{assump: comparison of the ad and af channel values}, and Lemma \ref{lem : concavity}, one can conclude that ${\rm MSE}^{\rm c}$ is no larger than ${\rm MSE}^{\rm af}$.
		 The proof is complete. \hfill $\blacksquare$

	\section{Proof of Lemma \ref{lem : concavity}}\label{appd : proof of lemma 2}
		Let's first consider the case of $K=2$. In this case, we have $a_1+a_2 = b_1+b_2$, and $a_1 \leq b_1, b_2 \leq a_2$. By denoting $\delta = b_1-a_1 = a_2-b_2 \geq 0$, we have
		\begin{subequations}
			\begin{align}
				&f(b_1) + f(b_2) - f(a_1) -f(a_2) \\
				=& f(a_1 +\delta) - f(a_1) - (f(b_2+\delta) - f(b_2))\\
				\circeq& \lim_{\delta \rightarrow 0^+}\frac{f(a_1 +\delta) - f(a_1)}{\delta} - \frac{f(b_2+\delta) - f(b_2)}{\delta} \label{eqn: mse fucntion iequality 1}\\
				=& f'(a_1) - f'(b_2) \geq 0, \label{eqn: mse fucntion iequality 2}
			\end{align}
		\end{subequations}
		where $x \circeq y$ signifies that $x$ and $y$ have the same sign. \eqref{eqn: mse fucntion iequality 1} is true since $f'(\cdot) > 0$, and the inequality in \eqref{eqn: mse fucntion iequality 2} is due to the fact that $f''(\cdot) < 0$. Moreover, the equality in \eqref{eqn: mse fucntion iequality 2} holds only when $a_1=b_1$ and $a_2=b_2$.
		
		For the case of $K>2$, let $\delta_k \triangleq b_k - a_k, k=1,...,K$. It is evident that $\sum_{k=1}^K \delta_k = 0$. For ease of analysis, we rearrange $\{\delta_k\}_{k=1}^K$ in the descending order and denote it by 
		\begin{align}
			\{\Delta_\ell\}_{\ell=1}^K = \{\Delta_{\ell} | \Delta_{\ell} \geq \Delta_{\ell+1}, \Delta_\ell \in  \{\delta_k\}_{k=1}^K, \ell =1,...,K\}.
		\end{align}
		It is ready to see that 
		\begin{subequations}
			\begin{align}
				&\Delta_1 = \max_k\{\delta_k\} \geq b_1-a_1 \geq 0,\\
				&\Delta_K = \min_k\{\delta_k\} \leq b_K-a_K \leq 0, \\
				&\sum_{k=1}^K \Delta_k = 0, \label{eqn: Delta summation property1}\\
				&\sum_{k=1}^{\bar K} \Delta_k  \geq 0,\forall 1\leq \bar K \leq K,\label{eqn: Delta summation property2}
			\end{align}
		\end{subequations}
		where \eqref{eqn: Delta summation property1} is due to $\sum_{k=1}^K a_k = \sum_{k=1}^K b_k$ and \eqref{eqn: Delta summation property2} is due to that  $\{\Delta_\ell\}_{\ell=1}^K$ are arranged in the descending order. 
		Meanwhile, rearrange $\{a_k\}_{k=1}^K$ and $\{b_k\}_{k=1}^K$ accordingly and denote them by
		\begin{align}
			\begin{cases}
				c_\ell = a_k \\
				d_\ell = b_k
			\end{cases},
			{\rm if}~ \Delta_\ell = \delta_k = b_k-a_k, \ell,k=1,...,K.
		\end{align}
		Thus, $\Delta_\ell = d_\ell- c_\ell, \ell = 1,...,K$.
		With the above definitions, we equivalently prove $\sum_{\ell=1}^K f(c_{\ell}) \leq \sum_{\ell=1}^K f(d_{\ell})$. 
		
		Note that $\Delta_1 = \max\{\delta_k\} \geq 0$. Then, according to the conclusion in the case of $K=2$, we have
		\begin{align}\label{eqn: mse fucntion inequality N1}
			f(d_1-\Delta_1) + f(d_2+\Delta_1) \leq f(d_1) + f(d_2)
		\end{align}
		Now, let $\widetilde{d}_2 \triangleq d_2 + \Delta_1$, we have $\widetilde{d}_2 \geq c_2$ since $\Delta_1 + \Delta_2 \geq 0$. Denote $\widetilde \Delta_2 = \widetilde d_2 - c_2 = \Delta_1 + \Delta_2 \geq 0$, again we have
		\begin{align}\label{eqn: mse fucntion inequality N2}
			f(\widetilde d_2-\widetilde\Delta_2) + f(d_3+\widetilde\Delta_2) \leq f(\widetilde d_2) + f(d_3)
		\end{align}
		Then, for $\ell\geq 3$, let $\widetilde{d}_{\ell} \triangleq d_{\ell} + \widetilde\Delta_{{\ell}-1}$, we have $\widetilde{d}_{\ell} \geq c_{\ell}$. Analogously, by denoting $\widetilde\Delta_\ell = \widetilde d_{\ell} - c_{\ell} = \sum_{j=1}^{\ell} \Delta_{j} \geq 0$, we have
		\begin{align}\label{eqn: mse fucntion inequality N3}
			f(\widetilde d_\ell - \widetilde\Delta_\ell) + f(d_{\ell+1} + \widetilde\Delta_\ell) \leq f(\widetilde d_\ell) + f(d_{\ell+1}),  \ell=3,...,K-1.
		\end{align}
		Taking summation over the two sides of \eqref{eqn: mse fucntion inequality N1} \eqref{eqn: mse fucntion inequality N2} and \eqref{eqn: mse fucntion inequality N3} together, we arrive at
		\begin{align}\label{eqn: concavity summation 1}
			f(d_1-\Delta_1) + \sum_{\ell=2}^{K-1} f(\widetilde d_\ell - \widetilde\Delta_\ell) + \sum_{\ell=1}^{K-1} f(d_{\ell+1}+\widetilde\Delta_\ell)
			\leq  \sum_{\ell=2}^{K-1}f(\widetilde d_\ell) + \sum_{\ell=1}^{K} f(d_\ell), 
		\end{align} 
		Since $d_1-\Delta_1 = c_1$, $c_\ell = \widetilde d_\ell - \widetilde\Delta_\ell$, and $\widetilde{d}_{\ell} = d_{\ell} + \widetilde\Delta_{{\ell}-1}$, \eqref{eqn: concavity summation 1} can be written as 
		\begin{align}
			\sum_{\ell=1}^{K-1} f(c_k) + \sum_{\ell=2}^{K}f(\widetilde{d}_\ell) \leq \sum_{\ell=2}^{K-1}f(\widetilde d_\ell) + \sum_{\ell=1}^{K} f(d_\ell)
		\end{align}
		and thus
		\begin{align}
			\sum_{\ell=1}^{K-1} f(c_k) + f(\widetilde{d}_K) \leq \sum_{\ell=1}^{K} f(b_\ell)
		\end{align}
		Since $\widetilde\Delta_K = \sum_{\ell=1}^K \Delta_\ell = 0$ by \eqref{eqn: Delta summation property1}, we have $f(\widetilde d_K) = f(c_K)$. Therefore, we conclude that $\sum_{k=1}^{K} f(c_k) \leq \sum_{k=1}^{K} f(d_k)$, and $\sum_{k=1}^K f(a_k) \leq \sum_{k=1}^K f(b_k)$. This completes the proof of Lemma \ref{lem : concavity}. \hfill $\blacksquare$

	\section{Proof of Proposition \ref{prop : nmse aggregation}}\label{appd: proof of mse of age algorithm}
		Since 
		 $[\widehat{\Gb}_m]_j$, 
		$[\widehat{\Gb}^{(l)}_m]_j$, and 
		$\alpha [\widehat{\Gb}^{(l)}_m]_j + (1\!-\!\alpha) [\widehat{\Gb}^{(l)}_m]_j$, 
		are determined independently, one can derive their corresponding MSEs separately. The MSE for centrally estimating $[\widehat{\Gb}]_j$ is given by \eqref{eqn: mse central age algorithm}.
		For the locally estimated $[\widehat{\Gb}^{(l)}_m]_j, m \in \Mset$, similar to \eqref{eqn: mse angle delay domain channel local}, the corresponding MSEs is given by 
			\begin{align}
				{\rm MSE}_j^{\rm FD} &= \sum_{m=1}^M \sum_{i=1}^{N_r}  \frac{[\Rb_{\Hbs_m}]_{i,j} \sigma_w^2 }{[\Rb_{\Hbs_m}]_{i,j} + \sigma_w^2}, j \in \bar\Ic_m^{(l)}.
			\end{align}
		While, for $[\widetilde{\Ab}_m]_j, j \in \Ic\cap\Ic_m^{(l)}$, which is determined by convexly combing $[\widehat{\Gb}_{m}]_j$ and $[\widehat{\Gb}^{(l)}_{m}]_j$. Thus, the corresponding MSE is given by 
			\begin{align}
				&\sum_{m=1}^{M}\left\|\alpha [\widehat{\Gb}_m]_j \!+\! (1\!-\!\alpha) [\widehat{\Gb}^{(l)}_m]_j - [\widetilde\Hb_m]_j \right\|^2 \notag\\
				&\leq\! \alpha \!\sum_{m=1}^{M}\!\left\|[\widehat{\Gb}_m]_j \!-\! [\widetilde\Hb_m]_j \right\|^2 \!\!+\! (1\!-\!\alpha)\! \sum_{m=1}^{M}\! \left\| [\widehat{\Gb}^{(l)}_m]_j \!-\! [\widetilde\Hb_m]_j \right\|^2 \notag\\
				& = \alpha {\rm MSE}_{j}^{\rm c,AGE} + (1-\alpha){\rm MSE}_{j}^{\rm FD}
			\end{align}
		where $\widetilde\Hb_m$ is the antenna-and-delay-domain channel of node $m$. Summarily, the MSE of the AGE-based algorithm is upper-bounded by {\smaller $\sum_{j \in \bar\Ic_m} \!\!{\rm MSE}_j^{\rm c,AGE} \!+\! \sum_{j \in \bar\Ic_m^{(l)}}\!\!{\rm MSE}_j^{\rm FD} \!+\! \sum_{j \in \Ic\cap\Ic_m^{(l)}}\!\left(\!\alpha {\rm MSE}_j^{\rm c,AGE} \!+\! (1\!-\!\alpha){\rm MSE}_j^{\rm FD}\!\right)$}. Since {\small ${\rm MSE}_j^{\rm c,AGE} \!\leq\! {\rm MSE}_j^{\rm FD},\forall j \in \bar\Ic_m^{(l)}$}, one can conclude that {\small ${\rm MSE}^{\rm c,AGE} \leq {\rm MSE}^{\rm FD}$} always hold. The proof is complete. \hfill $\blacksquare$

\end{appendices}

	% The very first letter is a 2 line initial drop letter followed
	% by the rest of the first word in caps.
	%
	% form to use if the first word consists of a single letter:
	% \IEEEPARstart{A}{demo} file is ....
	%
	% form to use if you need the single drop letter followed by
	% normal text (unknown if ever used by the IEEE):
	% \IEEEPARstart{A}{}demo file is ....
	%
	% Some journals put the first two words in caps:
	% \IEEEPARstart{T}{his demo} file is ....
	%
	% Here we have the typical use of a "T" for an initial drop letter
	% and "HIS" in caps to complete the first word.
	
%	
% \subsection{Related Works}
%
%  There are still many technical challenges to overcome in order to harvest the benefits of UAV-enabled wireless communications {\blue \cite{Tutorial,Zy_uavTutorial}.}
%
%
%	\subsection{Contributions}
%	
%	In this paper,
%	

%
%	\begin{enumerate}
%		
%		\item
%
%	\end{enumerate}
%	
%
%
%	
%	\vspace{-0.1cm}
%	\section{System Model and Problem Formulation} \label{sec:sytem_model}	\vspace{-0.15cm}
%
%	\subsection{System Model} \label{subsec:sytem_model}
%
%	
%
%	
%	\vspace{-0.32cm}
%	\section{Conclusion \label{sec:conclusion}}

	% if have a single appendix:
	%\appendix[Proof of the Zonklar Equations]
	% or
	%\appendix  % for no appendix heading
	% do not use \section anymore after \appendix, only \section*
	% is possibly needed
	
	% use appendices with more than one appendix
	% then use \section to start each appendix
	% you must declare a \section before using any
	% \subsection or using \label (\appendices by itself
	% starts a section numbered zero.)
	%
	
%	\newpage
%	\appendices
%	\vspace{-0.32cm}	

\smaller[1]

\begin{thebibliography}{10}
	\providecommand{\url}[1]{#1}
	\csname url@samestyle\endcsname
	\providecommand{\newblock}{\relax}
	\providecommand{\bibinfo}[2]{#2}
	\providecommand{\BIBentrySTDinterwordspacing}{\spaceskip=0pt\relax}
	\providecommand{\BIBentryALTinterwordstretchfactor}{4}
	\providecommand{\BIBentryALTinterwordspacing}{\spaceskip=\fontdimen2\font plus
		\BIBentryALTinterwordstretchfactor\fontdimen3\font minus
		\fontdimen4\font\relax}
	\providecommand{\BIBforeignlanguage}[2]{{%
			\expandafter\ifx\csname l@#1\endcsname\relax
			\typeout{** WARNING: IEEEtran.bst: No hyphenation pattern has been}%
			\typeout{** loaded for the language `#1'. Using the pattern for}%
			\typeout{** the default language instead.}%
			\else
			\language=\csname l@#1\endcsname
			\fi
			#2}}
	\providecommand{\BIBdecl}{\relax}
	\BIBdecl
	
	\bibitem{Larsson2014CM}
	E.~G. Larsson, O.~Edfors, F.~Tufvesson, and T.~L. Marzetta, ``Massive {MIMO}
	for next generation wireless systems,'' \emph{IEEE Commun. Mag.}, vol.~52,
	no.~2, pp. 186--195, Feb. 2014.
	
	\bibitem{Lu2014JSTSP}
	L.~Lu, G.~Y. Li, A.~L. Swindlehurst, A.~Ashikhmin, and R.~Zhang, ``An overview
	of massive {MIMO}: Benefits and challenges,'' \emph{IEEE J. Sel. Top. Signal
		Process.}, vol.~8, no.~5, pp. 742--758, May 2014.
	
	\bibitem{Gershman2010SPM}
	A.~B. Gershman, N.~D. Sidiropoulos, S.~Shahbazpanahi, M.~Bengtsson, and
	B.~Ottersten, ``Convex optimization-based beamforming,'' \emph{IEEE Signal
		Process Mag.}, vol.~27, no.~3, pp. 62--75, May 2010.
	
	\bibitem{Luo2010SPM}
	Z.-Q. Luo, W.-K. Ma, A.~M.-C. So, Y.~Ye, and S.~Zhang, ``Semidefinite
	relaxation of quadratic optimization problems,'' \emph{IEEE Signal Process
		Mag.}, vol.~27, no.~3, pp. 20--34, May 2010.
	
	\bibitem{Li2017TSP}
	Y.~Li, C.~Tao, G.~Seco-Granados, A.~Mezghani, A.~L. Swindlehurst, and L.~Liu,
	``Channel estimation and performance analysis of one-bit massive {MIMO}
	systems,'' \emph{IEEE Trans. Signal Process.}, vol.~65, no.~15, pp.
	4075--4089, Aug. 2017.
	
	\bibitem{WangTSP2018}
	B.~Wang, F.~Gao, S.~Jin, H.~Lin, and G.~Y. Li, ``Spatial- and
	frequency-wideband effects in millimeter-wave massive {MIMO} systems,''
	\emph{IEEE Trans. Signal Process.}, vol.~66, no.~13, pp. 3393--3406, Jul.
	2018.
	
	\bibitem{Yang2001TCOM}
	B.~Yang, Z.~Cao, and K.~Letaief, ``Analysis of low-complexity windowed
	{DFT}-based {MMSE} channel estimator for {OFDM} systems,'' \emph{IEEE Trans.
		Commun.}, vol.~49, no.~11, pp. 1977--1987, Nov. 2001.
	
	\bibitem{chang-tsp2010}
	T.-H. Chang, W.-C. Chiang, Y.-W.~P. Hong, and C.-Y. Chi, ``Training sequence
	design for discriminatory channel estimation in wireless {MIMO} systems,''
	\emph{IEEE Trans. Signal Process.}, vol.~58, no.~12, pp. 6223--6237, Dec.
	2010.
	
	\bibitem{Takano2018TWC}
	Y.~Takano, H.-J. Su, M.~Juntti, and T.~Matsumoto, ``A conditional ${\ell}1$
	regularized {MMSE} channel estimation technique for {IBI} channels,''
	\emph{IEEE Trans. Wireless Commun.}, vol.~17, no.~10, pp. 6720--6734, Oct.
	2018.
	
	\bibitem{Emil2010TSP}
	E.~Bj\"{o}rnson and B.~Ottersten, ``A framework for training-based estimation
	in arbitrarily correlated rician {MIMO} channels with rician disturbance,''
	\emph{IEEE Trans. Signal Process.}, vol.~58, no.~3, pp. 1807--1820, Mar.
	2010.
	
	\bibitem{bjornson-jstsp2014}
	N.~Shariati, E.~Bj\"{o}rnson, M.~Bengtsson, and M.~Debbah, ``Low-complexity
	polynomial channel estimation in large-scale {MIMO} with arbitrary
	statistics,'' \emph{IEEE J. Sel. Topics Signal Process.}, vol.~8, no.~5, pp.
	815--830, Oct. 2014.
	
	\bibitem{Gao2020TWC}
	S.~Gao, X.~Cheng, and L.~Yang, ``Estimating doubly-selective channels for
	hybrid mmwave massive {MIMO} systems: A doubly-sparse approach,'' \emph{IEEE
		Trans. Wireless Commun.}, vol.~19, no.~9, pp. 5703--5715, Sept. 2020.
	
	\bibitem{Fan2018TWC}
	D.~Fan, F.~Gao, Y.~Liu, Y.~Deng, G.~Wang, Z.~Zhong, and A.~Nallanathan, ``Angle
	domain channel estimation in hybrid millimeter wave massive {MIMO} systems,''
	\emph{IEEE Trans. Wireless Commun.}, vol.~17, no.~12, pp. 8165--8179, Dec.
	2018.
	
	\bibitem{Kim2019TCOM}
	H.~Kim, G.-T. Gil, and Y.~H. Lee, ``Two-step approach to time-domain channel
	estimation for wideband millimeter wave systems with hybrid architecture,''
	\emph{IEEE Trans. Commun.}, vol.~67, no.~7, pp. 5139--5152, Jul. 2019.
	
	\bibitem{Masood2015TSP}
	M.~Masood, L.~H. Afify, and T.~Y. Al-Naffouri, ``Efficient coordinated recovery
	of sparse channels in massive {MIMO},'' \emph{IEEE Trans. Signal Process.},
	vol.~63, no.~1, pp. 104--118, Jan. 2015.
	
	\bibitem{gaoTSP2015}
	Z.~Gao, L.~Dai, Z.~Wang, and S.~Chen, ``Spatially common sparsity based
	adaptive channel estimation and feedback for {FDD} massive {MIMO},''
	\emph{IEEE Trans. Signal Process.}, vol.~63, no.~23, pp. 6169--6183, Dec.
	2015.
	
	\bibitem{Li2017JESTCS}
	K.~Li, R.~R. Sharan, Y.~Chen, T.~Goldstein, J.~R. Cavallaro, and C.~Studer,
	``Decentralized baseband processing for massive {MU-MIMO} systems,''
	\emph{IEEE J. Emerging Sel. Top. Circuits Syst.}, vol.~7, no.~4, pp.
	491--507, Dec. 2017.
	
	\bibitem{Jeon2019TSP}
	C.~Jeon, K.~Li, J.~R. Cavallaro, and C.~Studer, ``Decentralized equalization
	with feedforward architectures for massive {MU-MIMO},'' \emph{IEEE Trans.
		Signal Process.}, vol.~67, no.~17, pp. 4418--4432, Sept. 2019.
	
	\bibitem{Jeon2017ISIT}
	------, ``On the achievable rates of decentralized equalization in massive
	{MU-MIMO} systems,'' in \emph{IEEE Int. Symp. Inf. Theory}, Jun. 2017, pp.
	1102--1106.
	
	\bibitem{Zhang2020TVT}
	Z.~Zhang, H.~Li, Y.~Dong, X.~Wang, and X.~Dai, ``Decentralized signal detection
	via expectation propagation algorithm for uplink massive {MIMO} systems,''
	\emph{IEEE Trans. Veh. Technol.}, vol.~69, no.~10, pp. 11\,233--11\,240, Oct.
	2020.
	
	\bibitem{Amiri2022WCL}
	A.~Amiri, C.~N. Manchón, and E.~de~Carvalho, ``Uncoordinated and decentralized
	processing in extra-large {MIMO} arrays,'' \emph{IEEE Wireless Commun.
		Lett.}, vol.~11, no.~1, pp. 81--85, Jan. 2022.
	
	\bibitem{Rusek-TSP2020}
	J.~Rodr\'iguez~S\'anchez, F.~R\'usek, O.~Edfors, M.~Sarajli\'c, and L.~Liu,
	``Decentralized massive {MIMO} processing exploring daisy-chain architecture
	and recursive algorithms,'' \emph{IEEE Trans. Signal Process.}, vol.~68, pp.
	687--700, Jan. 2020.
	
	\bibitem{Rusek-TSP2021}
	J.~V. Alegr\'ia, F.~R\'usek, and O.~Edfors, ``Trade-offs in decentralized
	multi-antenna architectures: The {WAX} decomposition,'' \emph{IEEE Trans.
		Signal Process.}, vol.~69, pp. 3627--3641, Jun. 2021.
	
	\bibitem{Rusek-TSP2022}
	J.~Rodr\'iguez~S\'anchez, F.~R\'usek, O.~Edfors, and L.~Liu, ``Distributed and
	scalable uplink processing for {LIS}: Algorithm, architecture, and design
	trade-offs,'' \emph{IEEE Trans. Signal Process.}, vol.~70, pp. 2639--2653,
	Apr. 2022.
	
	\bibitem{Zhang2022TVT}
	Z.~Zhang, Y.~Dong, K.~Long, X.~Wang, and X.~Dai, ``Decentralized baseband
	processing with gaussian message passing detection for uplink massive
	{MU-MIMO} systems,'' \emph{IEEE Trans. Veh. Technol.}, vol.~71, no.~2, pp.
	2152--2157, Feb. 2022.
	
	\bibitem{Li2019CSSC}
	K.~Li, J.~McNaney, C.~Tarver, O.~Castañeda, C.~Jeon, J.~R. Cavallaro, and
	C.~Studer, ``Design trade-offs for decentralized baseband processing in
	massive {MU-MIMO} systems,'' in \emph{53rd Asilomar Conf. Signals, Syst.,
		Comput.}, Nov. 2019, pp. 906--912.
	
	\bibitem{Muris-WCL2019}
	M.~Sarajli\'c, F.~R\'usek, J.~Rodríguez~S\'anchez, L.~Liu, and O.~Edfors,
	``Fully decentralized approximate zero-forcing precoding for massive mimo
	systems,'' \emph{IEEE Wireless Commun. Lett.}, vol.~8, no.~3, pp. 773--776,
	Jun. 2019.
	
	\bibitem{Li2018ACSSC}
	K.~Li, C.~Jeon, J.~R. Cavallaro, and C.~Studer, ``Feedforward architectures for
	decentralized precoding in massive {MU-MIMO} systems,'' in \emph{Asilomar
		Conf. Signals, Syst., Comput.}, Oct. 2018, pp. 1659--1665.
	
	\bibitem{Croisfelt-ACSSC2021}
	V.~Croisfelt, T.~Abr\~ao, A.~Amiri, E.~de~Carvalho, and P.~Popovski,
	``Decentralized design of fast iterative receivers for massive {MIMO} with
	spatial non-stationarities,'' in \emph{55rd Asilomar Conf. Signals, Syst.,
		Comput.}, 2021, pp. 1242--1249.
	
	\bibitem{Kulkarni2021TCS}
	A.~Kulkarni, M.~A. Ouameur, and D.~Massicotte, ``Hardware topologies for
	decentralized large-scale {MIMO} detection using newton method,'' \emph{IEEE
		Trans. Circuits Syst. I Regul. Pap.}, vol.~68, no.~9, pp. 3732--3745, Sept.
	2021.
	
	\bibitem{AlamTCOM2016}
	A.~Zaib, M.~Masood, A.~Ali, W.~Xu, and T.~Y. Al-Naffouri, ``Distributed channel
	estimation and pilot contamination analysis for massive {MIMO-OFDM}
	systems,'' \emph{IEEE Trans. Commun.}, vol.~64, no.~11, pp. 4607--4621, Nov.
	2016.
	
	\bibitem{quadriga}
	J.~Stephan, R.~Leszek, B.~Kai, and T.~Lars, ``{QuaDRiGa}: A 3-{D} multi-cell
	channel model with time evolution for enabling virtual field trials,''
	\emph{IEEE Trans. Antennas Propagat.}, vol.~62, no.~6, pp. 3242--3256, 2014.
	
	\bibitem{Gong2017TVT}
	B.~Gong, L.~Gui, Q.~Qin, X.~Ren, and W.~Chen, ``Block distributed compressive
	sensing-based doubly selective channel estimation and pilot design for
	large-scale {MIMO} systems,'' \emph{IEEE Trans. Veh. Technol.}, vol.~66,
	no.~10, pp. 9149--9161, Oct. 2017.
	
\end{thebibliography}
\end{document}